\documentstyle[12pt,a4,epsf]{article}

\newcommand{\eqpt}{\hspace{6pt}.}           
\newcommand{\eqcm}{\hspace{6pt},}

\newcommand{\eqref}[1]{(\ref{#1})}          

\newcommand{\arreq}{\hspace{0.4em} = \hspace{0.4em}}
\newcommand{\lsim}{\raisebox{-.5ex}{\footnotesize$
     \,\:\stackrel{\textstyle<}{\sim}\,\:$}}    

\newcommand{\limit}[2]{\raisebox{-1.5ex}{ $\stackrel{\textstyle
      \longrightarrow}{\scriptstyle #1 \rightarrow #2}$ }}

\newcommand{\pdot}{\! \cdot \!}             

\newcommand{\gp}{\gamma^{\ast} p}           
\newcommand{\eps}{\varepsilon}              
\newcommand{\pom}{{I \! \! P}}              

\newcommand{\vp}{h}                         
\newcommand{\vpb}{{\bf h}}

\newcommand{\Mx}{M_X}                       
\newcommand{\mq}{m_q}
\newcommand{\dq}{\chi}                      
\newcommand{\qq}{P}
\newcommand{\Pl}{{P}_L}
\newcommand{\Pt}{{\bf P}_T}
\newcommand{\Plcut}{P_{L {\it cut}}}
\newcommand{\Ptcut}{{\bf P}_{T {\it cut}}}
\newcommand{\phiqq}{\varphi_{q \bar{q}}}
\newcommand{\qj}{P_{\it FB}}
\newcommand{\Pfb}{| \Pl |}
\newcommand{\phifb}{\varphi_{\it FB}}
\newcommand{\lt}{{\bf l}_T} 
\newcommand{\dt}{{\bf \Delta}_T} 
\newcommand{\dl}{\delta}

\newcommand{\phiX}{\varphi_X}

\newcommand{\iv}{v}
\newcommand{\iw}{w}
\newcommand{\ib}{b}
\newcommand{\sm}{a}

\newcommand{\PlX}{P_L^\ast}                 
\newcommand{\PlPlX}{P_L^{\ast 2}} 
\newcommand{\PtX}{{\bf P}_T^\ast}
\newcommand{\PtPtX}{{\bf P}_T^{\ast 2}}
\newcommand{\PtXcut}{{\bf P}_{T {\it cut}}^{\ast 2}} 
\newcommand{\pp}{\chi^\ast}

\newcommand{\ibX}{b^\ast}
\newcommand{\iwX}{w^\ast}
\newcommand{\iwiwX}{w^{\ast 2}}
\newcommand{\smX}{a^\ast}
\newcommand{\smsmX}{a^{\ast 2}}
\newcommand{\scaleX}{\lambda^{\ast 2}}

\newcommand{\GeV}{\mbox{\ GeV}}             

\begin{document}

\begin{flushright}
{CPTH-S472-1096 \\
  DAMTP-96-91}
\end{flushright}
\vspace{2\baselineskip}
\begin{center}
{\bf\LARGE AZIMUTHAL ANGLES \\
\bigskip
IN DIFFRACTIVE $ep$ COLLISIONS \\}
\vspace{3\baselineskip}
M. Diehl\footnote{email: diehl@orphee.polytechnique.fr} \\
\vspace{\baselineskip}
Centre de Physique Th\'eorique\footnote{Unit\'e propre 14 du CNRS}\\
Ecole Polytechnique\\ 91128 Palaiseau Cedex, France\\
and\\
Department of Applied Mathematics and Theoretical Physics\\ University
of Cambridge\\ Cambridge CB3 9EW, England
\vfill
\begin{abstract}
  We investigate azimuthal correlations in deep inelastic diffractive
  scattering, $e+p \to e+\tilde{p}+X$. The dependence of the $ep$
  cross section on the angle between the lepton plane and some
  direction in the hadronic final state can be written in a simple
  form; its measurement can be used to constrain the cross section for
  longitudinally polarised photons. Using the model of
  nonperturbative two-gluon exchange of Landshoff and Nachtmann we
  calculate the distribution of the azimuthal jet angle in diffractive
  dijet production and find that useful bounds on the longitudinal
  cross section for such events might be obtained from its
  measurement. We then discuss the predictions of this model for the
  dependence of the $ep$ cross section on the azimuthal angle of the
  proton remnant $\tilde{p}$, which contains information about the
  helicity content of the pomeron.
\end{abstract}
\end{center}

\newpage

\section{Introduction}
\label{sec:intro}

Our knowledge of diffractive physics in deep inelastic electron-proton
scattering is greatly increasing as diffractive events are being
studied in more and more detail at HERA \cite{HERA}. The phenomenology
of these events has many aspects, and several theoretical models have
been proposed to describe them
\cite{IS,NikZak,BarLottWust,Dispersion,MD,BuchHeb,Ingelman}. Despite
various successes of these models we are yet far from a clear
theoretical picture of what pomeron physics is in terms of QCD.
Detailed studies of the characteristics of the final state might help
to further our understanding of the underlying mechanisms and to
distinguish between various models.

The measurement of two different kinds of azimuthal angles has
recently been proposed: the azimuthal angle of the scattered proton
\cite{HDCB,GehrStir} or, equivalently, of the diffractive system as a
whole, and the azimuthal angle of the jets in events with only two
jets of large transverse momentum in the diffractive final state
\cite{BarLott}. The present paper will be concerned with both issues
and has two purposes: to discuss some general aspects of azimuthal
distributions in diffraction and to present in detail predictions for
such distributions in the model of nonperturbative two-gluon exchange
of Landshoff and Nachtmann \cite{LN}.

The structure of this paper is as follows. In sec.~\ref{sec:single} we
generalise the formalism of \cite{HDCB} for azimuthal distributions in
diffraction and show which constraints on the $\gp$ cross section for
longitudinal photons their measurement can provide. A corresponding
framework has long been used in various processes in non-diffractive
DIS \cite{OrdinaryDIS,Zerwas}. As an application we consider in
sec.~\ref{sec:jets} the azimuthal angle of the jets in diffractive
dijet production. Some features of its distribution are quite
characteristic for two-gluon exchange and might offer a way to test
the two-gluon approximation in this type of events as was already
pointed out in \cite{BarLott}. We calculate the angular dependence in
the Landshoff-Nachtmann model and show which bounds on the
longitudinal cross sections could be obtained from its measurement. We
also show how this method can be generalised to final states that do
not necessarily have two-jet topology. In sec.~\ref{sec:finite} we
generalise the calculation to nonzero $t$ and obtain the corrections
this gives for the $\gp$ cross sections and for the distribution of
the azimuthal jet angle. Using this calculation we investigate a
genuine finite-$t$ effect in sec.~\ref{sec:proton}: the correlation
between the azimuthal angles of the scattered lepton and proton. In
\cite{HDCB} it was shown that this observable contains information
about the helicity structure of the pomeron and argued that it might
provide a sensitive test of various theoretical ideas about the
underlying dynamics. We conclude with a summary in sec.~\ref{sec:sum}.

\section{Azimuthal angle dependence in diffraction}
\label{sec:single}
\setcounter{equation}{0}

We begin by extending the formalism of \cite{HDCB} to a large class of
azimuthal angles in diffractive electron-proton or positron-proton
collisions,
\newpage
\begin{equation}
e(k) + p(p) \to e(k') + \tilde{p}(\tilde{p}) + X(p_X) \eqcm
  \label{DiffReac}
\end{equation}
where the proton remnant $\tilde{p}$ can be a proton or a diffractive
excitation of a proton and where four-momenta are indicated in
parentheses. We will use the conventional kinematic quantities $Q^2,
W^2, x, y, s, t$ for deep inelastic scattering, $\Mx$ for the
invariant mass of the diffractive system $X$, and the variables $\beta
= Q^2 / (Q^2 + \Mx^2 - t)$ and $\xi = (Q^2 + \Mx^2 - t) /(W^2 + Q^2 -
m_p^2)$.

Working in the $\gp$ rest frame one can write the azimuthal dependence
of the $e p$ cross section in a simple way by making use of the
factorisation of \eqref{DiffReac} into $\gamma^\ast$ emission by the
electron or positron and a diffractive photon-proton collision
\begin{equation}
\gamma^\ast(q) + p(p) \to  X(p_X) + \tilde{p}(\tilde{p}) \eqpt
  \label{PhotReac}
\end{equation}
To achieve this it is essential that the selection of diffractive
events, e.g.\ the definition of a rapidity gap between $X$ and
$\tilde{p}$, is unaffected by a common rotation about the $\gp$ axis
of the momenta in the hadronic final state $X \tilde{p}$, with the
lepton momenta $k$ and $k'$ being kept fixed. This is guaranteed if
the selection criteria only involve quantities of the $\gp$ reaction,
i.e.\ if they do {\em not\/} refer to the lepton momenta $k$ and $k'$.
Examples for such criteria have been given in \cite{HDCB}.

We define an azimuthal angle with respect to a direction in the
hadronic final state $X \tilde{p}$. To this end we introduce a
four-vector $\vp$ which depends only on particle momenta in the $\gp$
reaction \eqref{PhotReac}, i.e.\ on $p$, $q$ and the momenta of the
final state hadrons. Using a right-handed Cartesian coordinate system
with the $z$ axis in the direction of the photon momentum ${\bf q}$
and some fixed $x$ and $y$ axes we define $\varphi$ as the azimuthal
angle between the lepton momentum ${\bf k}$ and the vector $\vpb$,
i.e.\ as the azimuthal angle of ${\bf k}$ minus the azimuthal angle of
$\vpb$.

We also use $\vp$ to introduce polarisation vectors for the virtual
photon:
\begin{eqnarray}
\eps_0 & = & \frac{1}{Q \sqrt{1 + m_p^2 \, Q^2 / (p \pdot
    q)^2}} \left( q + \frac{Q^2}{p \pdot q} \, p
    \right) \eqcm \hspace{1em}
\eps_1 \arreq \frac{\vp_T}{\sqrt{- \vp_T^2}} \eqcm \hspace{1em}
\eps_2 \arreq \frac{n}{\sqrt{- {n}^2}}  \eqcm \nonumber \\
  \label{polaris}
\end{eqnarray}
where
\begin{eqnarray}
  \vp_T = \vp - \frac{(p \pdot q) (p \pdot \vp) - m_p^2 \, (q \pdot
    \vp)}{(p \pdot q)^2 + m_p^2 \, Q^2} \, q - \frac{(p \pdot q) (q
    \pdot \vp) + Q^2 (p \pdot \vp)}{(p \pdot q)^2 + m_p^2 \, Q^2} \, p
  \label{eps1}
\end{eqnarray}
is the transverse part of $\vp$ with respect to $p$ and $q$, and $
{n}^{\mu} = \eps^{\mu \nu \rho \sigma} \, p_{\nu} q_{\rho}
\vp_{\sigma}$ is normal to $p$, $q$ and $\vp$. Polarisations for
positive or negative photon helicity are as usual given by $
\eps_{\pm} = \mp (\eps_1 \pm i \eps_2) / \sqrt{2}$.

The contractions of $\eps_{-}^\mu$, $\eps_{0}^\mu$, $\eps_{+}^\mu$
with the appropriate matrix element of the hadronic electromagnetic
current $e J_\mu$ give the amplitudes $e {\cal M}_m$ for subreaction
\eqref{PhotReac} with photon helicity $m$,
\begin{equation}
  e {\cal M}_m = \langle X \tilde{p} {\rm \ out}
  \, | \, e J_\mu(0) \, | \, p \rangle \cdot \eps_m^\mu  \eqcm
  \hspace{3em}  m = -, 0, +  \eqpt
 \label{amplitude}
\end{equation}
The corresponding differential cross sections $d \sigma_{m m}$ are
obtained by multiplying ${\cal M}^{\ast}_{m} {\cal
  M}^{\phantom{\ast}}_{m}$ with the phase space element of the
hadronic final state $X \tilde{p}$ and with a normalisation factor,
summing over the states $X \tilde{p}$ allowed by our selection
criteria for the diffractive reaction and averaging over the initial
proton spin. Our normalisation factor corresponds to Hand's convention
\cite{Hand} for the photon flux. From ${\cal M}^{\ast}_{m} {\cal
  M}^{\phantom{\ast}}_{n}$ with $m \neq n$ we define in an analogous
manner differential interference terms $d \sigma_{m n}$ between
photons with helicities $m$ and $n$. It is easy to see that the matrix
$d \sigma_{m n}$ is hermitian, $d \sigma_{m n} = d \sigma^{\ast}_{n
  m}$.

With the requirement on the selection cuts formulated above the $d
\sigma_{m n}$ are invariant under a common rotation of the momenta in
the hadronic final state $X \tilde{p}$ about the $\gp$ axis. This is
because our transverse photon polarisations are not fixed but vary
with the final state as they depend on $\vp$. One can show that the
cross sections $d \sigma_{m m}$ are the same for different choices of
this vector, whereas the interference terms are not. In the following
sections we will put extra labels on the angle $\varphi$ and the $d
\sigma_{m n}$ to distinguish different choices of $\vp$, though for
the diagonals $d \sigma_{m m}$ this would not be necessary.

Integrating over the phase space of the hadronic final state we obtain
a matrix $\sigma_{m n}$. We will also consider $\gp$ cross sections
and interference terms that are differential in some kinematical
variables of the final state, such as the momentum of the proton
remnant or internal variables of the system $X$. We will only use
variables that can be defined as Lorentz invariant functions of the
four-momenta in the $\gp$ reaction. Provided that the selection
criteria for our reaction do not refer to any particular frame, the
differential cross sections and interference terms are then Lorentz
invariant and as a consequence depend only on $W^2$, $Q^2$ and the
variables in which they are differential.  Due to the rotation
invariance property just mentioned they are independent of the
azimuthal angle of $\vpb$ in our fixed coordinate system and hence
also of $\varphi$. An important property following from angular
momentum conservation is that interference terms which are
differential in the direction of $\vp$ vanish when $\vp$ becomes
collinear with $q$ and $p$, in which case the azimuthal angle
$\varphi$ is undefined \cite{HDCB}.

The $e p$ cross section can now be written as \cite{HDCB}
\begin{eqnarray}
\frac{d \sigma(e p \to e \tilde{p} X)}{d x \, d Q^2 \, d
  \varphi} &=& \frac{2 \tilde{\Gamma}}{2 \pi} \left\{ \frac{1}{2}
  (\sigma_{++}+\sigma_{--}) + \eps \sigma_{00} \right. \nonumber \\
&& \mbox{} - \eps\cos(2\varphi)\ {\rm Re}\, \sigma_{+-}
    + \eps\sin(2\varphi)\ {\rm Im}\, \sigma_{+-}
    \phantom{\sqrt{\eps(1+\eps)}}  \nonumber\\ 
&& \mbox{} - \sqrt{\eps(1+\eps)}\cos\varphi \ 
     {\rm Re}\, (\sigma_{+0} - \sigma_{-0})  \nonumber\\
&& \mbox{} + \sqrt{\eps(1+\eps)}\sin\varphi \
     {\rm Im}\, (\sigma_{+0} + \sigma_{-0})  \nonumber\\
&& \mbox{} + r_L\sqrt{1-\eps^2} \, \frac{1}{2}
     (\sigma_{++} - \sigma_{--})  \nonumber\\
&& \mbox{} - r_L\sqrt{\eps(1-\eps)}\cos\varphi \ {\rm Re}\, 
     (\sigma_{+0} + \sigma_{-0})  \nonumber\\
&& \left. \mbox{} + r_L\sqrt{\eps(1-\eps)}\sin\varphi \ {\rm Im}\, 
     (\sigma_{+0} - \sigma_{-0}) \right\}  \eqcm
  \label{XsectionMaster}
\end{eqnarray}
where we have integrated over a trivial overall angle, namely the
azimuthal angle of the scattered lepton in the $e p$ frame. $r_L$ is
the helicity of the incoming lepton, which is approximated to be
massless, $\eps = (1 - y)/ (1 - y + y^2 /2)$ is the usual ratio of
longitudinal and transverse photon flux and
\begin{equation}
  2 \tilde{\Gamma} = \frac{\alpha_{\it em}}{\pi Q^2} \, \frac{1-x}{x}
  \, \left( 1 - y + y^2 /2 \right)  \eqcm
  \label{flux} 
\end{equation}
where in the expressions of $\eps$ and $2 \tilde{\Gamma}$ we have
neglected terms of order $x^2 m_p^2 /Q^2$. Equation
\eqref{XsectionMaster} remains valid if its l.h.s.\ and the $\sigma_{m
  n}$ on its r.h.s.\ are made differential in additional variables as
described above. Since the $\gp$ cross sections and interference terms
are independent of $\varphi$ the dependence of the $e p$ cross section
on this angle is explicitly given by the trigonometric functions in
\eqref{XsectionMaster}.

Let us have a closer look at those combinations of the $\sigma_{m n}$
that are multiplied with the lepton helicity $r_L$ in
\eqref{XsectionMaster}. To make their role more apparent we introduce
differential cross sections and interference terms $d \sigma_{k l}$
with $k, l = 0, 1, 2$ analogous to the $d \sigma_{m n}$, but with the
linear photon polarisations $\eps_{0}$, $\eps_{1}$, $\eps_{2}$ of
\eqref{polaris} instead of $\eps_{-}$, $\eps_{0}$, $\eps_{+}$ in the
helicity basis. We have the relations
\begin{eqnarray}
\frac{1}{2} (d \sigma_{++}+d \sigma_{--}) &=& \frac{1}{2}
(d \sigma_{11} + d \sigma_{22}) \nonumber \\
{\rm Re}\, d \sigma_{+-} &=& - \frac{1}{2} (d \sigma_{11} -
d \sigma_{22}) \nonumber \\
{\rm Im}\, d \sigma_{+-} &=& {\rm Re}\, d \sigma_{12} \nonumber \\
{\rm Re}\, (d \sigma_{+0} - d \sigma_{-0}) &=& - \sqrt{2}\, {\rm Re}\,
d \sigma_{10}   \nonumber \\
{\rm Im}\, (d \sigma_{+0} + d \sigma_{-0}) &=& \sqrt{2}\,  {\rm Re}\,
d \sigma_{20} \nonumber \\
\frac{1}{2} (d \sigma_{++} - d \sigma_{--}) &=& - {\rm Im}\, d
\sigma_{12} \nonumber \\
{\rm Re}\, (d \sigma_{+0} + d \sigma_{-0}) &=& - \sqrt{2}\, {\rm Im}\,
d \sigma_{20} \nonumber \\
{\rm Im}\, (d \sigma_{+0} - d \sigma_{-0}) &=& - \sqrt{2}\, {\rm Im}\,
d \sigma_{10}  \eqpt
 \label{ToLinear}
\end{eqnarray}
The terms that depend on the lepton helicity in the $ep$ cross section
are seen to be the imaginary parts of $\gp$ interference terms for
linearly polarised photons. $d \sigma_{kl}$ is given by ${\cal
  M}_k^\ast {\cal M}_l^{\phantom{\ast}}$ multiplied with a phase space
element and real factors, summed over the appropriate final states and
averaged over the initial proton spin, where we define the amplitude
$e {\cal M}_k$ for the reaction $\gamma^\ast p \to X \tilde{p}$ with
photon polarisation $\eps_k$ in analogy to \eqref{amplitude}. The
imaginary parts of $d \sigma_{kl}$, $k \neq l$ are obviously zero if
the phase of this amplitude does not depend on the photon
polarisation.  Note that any (convention dependent) phase of $| \, X
\tilde{p} {\rm \ out} \rangle$ and $| \, p \rangle$ drops out in $d
\sigma_{kl}$.  Also there are no phases coming from $\eps_k^\ast$,
$\eps_l^{\phantom{\ast}}$ because for linearly polarised photons the
polarisation vectors are purely real; circular polarisation introduces
extra phases in the $\gp$ interference terms.

We emphasise that in order for ${\rm Im}\, d \sigma_{kl}$ to vanish
the ${\cal M}_k$ do {\em not\/} have to be real. The absence of final
state interactions gives vanishing interference terms if one sums over
a set of final states that is invariant under time reversal
(cf.~\cite{Zerwas}), but this is a sufficient condition, not a
necessary one. Here we are concerned with diffraction and the phases
of our amplitudes are certainly nonzero. However, for pure pomeron
exchange they are given by the signature factor and thus independent
of the $\gamma^\ast$ polarisation. It is beyond the scope of this
paper to investigate whether for example the superposition of pomeron
exchange with exchange of secondary trajectories or with multiple
exchanges could lead to polarisation dependent phases that might be
tested with longitudinally polarised lepton beams.

Going back to the $\sigma_{m n}$ for photons with definite helicity,
we now make use of the parity invariance of strong interactions. It
relates $\sigma_{m n}$ for different $m, n$, provided that the
selection criteria are parity invariant and that $\vp$ is a vector,
not a pseudovector. By an argument as in \cite{HDCB} one can show that
under these conditions
\begin{equation}
\sigma_{m n}(W^2,Q^2) = (-1)^{m+n} \, \sigma_{-m, -n}(W^2, Q^2)  \eqpt
  \label{ParityXsection}
\end{equation}
Using this and the hermiticity of $\sigma_{mn}$ one obtains the
relations
\begin{equation}
\sigma_{++} = \sigma_{--} , \hspace{2em}
\sigma_{+-} = \sigma_{-+} = \sigma_{+-}^{\ast} , \hspace{2em}
\sigma_{+0} = - \sigma_{-0} \eqcm
  \label{ParitySimpler}
\end{equation}
so that the expression \eqref{XsectionMaster} of the cross section
is simplified:
\begin{eqnarray}
\lefteqn{ \frac{d \sigma(e p \to e \tilde{p} X)}{d x \, d Q^2 \, d
  \varphi} = \frac{2 \tilde{\Gamma}}{2 \pi} \left\{\sigma_{++} +
  \eps \sigma_{00} - \eps\cos(2\varphi)\ \sigma_{+-}
  \phantom{\frac{1}{2}} \right. } \nonumber \\
& & \left. \mbox{} - 2 \sqrt{\eps(1+\eps)}\cos\varphi \ {\rm Re}\,
  \sigma_{+0}  + 2 r_L\sqrt{\eps(1-\eps)}\sin\varphi \ {\rm Im}\,
  \sigma_{+0}  \right\}  \eqpt
  \label{MasterSimpler}
\end{eqnarray}
For linear photon polarisations the relations corresponding to
\eqref{ParityXsection} read
\begin{equation}
\sigma_{20} = \sigma_{02} = \sigma_{21} = \sigma_{12} = 0  \eqcm
  \label{LinearZero}
\end{equation}
i.e.\ transverse photons with polarisation perpendicular to $\vp_T$ of
\eqref{eps1} do not interfere when the final state momenta are
integrated over. Expressions analogous to \eqref{ParityXsection} to
\eqref{LinearZero} are also valid for differential $\gp$ cross
sections and interference terms, provided that they depend only on
parity even variables, i.e.\ that one sums the $d \sigma_{m n}$ over a
parity invariant set of final states.

\subsection{Bounds on the cross section for longitudinal photons}
\label{sec:bounds}

We now show how the measurement of the $\varphi$-dependence in the
$ep$ cross section \eqref{MasterSimpler} can be used to constrain the
$\gp$ cross section for longitudinal photons as was pointed out in
\cite{HDCB}. $d \sigma_{m n}$ is a positive semidefinite matrix, which
with the simplifications from hermiticity and \eqref{ParityXsection}
from parity invariance implies \cite{HDCB}
\begin{equation}
\sigma_{++} + \sigma_{+-} \ge 0  \eqcm \hspace{4em}
\sigma_{00} (\sigma_{++} - \sigma_{+-}) \ge 2 |\sigma_{+0}|^2  \eqpt
  \label{eigenv}
\end{equation}
From the measurement of the $\varphi$-dependence in
\eqref{MasterSimpler} one can extract the weighted sum $\sigma_\eps =
\sigma_{++} + \eps \sigma_{00}$ of transverse and longitudinal $\gp$
cross sections as well as the interference terms $\sigma_{+-}$, ${\rm
  Re}\, \sigma_{+0}$ and ${\rm Im}\, \sigma_{+0}$. For ${\rm Im}\,
\sigma_{+0}$ one needs longitudinally polarised electron or positron
beams. With unpolarised beams one can use the weaker constraints
obtained by replacing $|\sigma_{+0}|$ with ${\rm Re}\, \sigma_{+0}$ in
\eqref{eigenv} and in the following.

Substituting $\sigma_\eps - \eps \sigma_{00}$ for $\sigma_{++}$ in
\eqref{eigenv} the first condition gives
\begin{equation}
\sigma_{00} \le \frac{\sigma_{\eps} + \sigma_{+-}}{\eps} \eqcm
  \label{constraint1}
\end{equation}
whereas the second becomes a quadratic inequality in $\sigma_{00}$
which leads to
\begin{eqnarray}
&& \frac{\sigma_{\eps} - \sigma_{+-}}{2 \eps} - \sqrt{
  \left(\frac{\sigma_{\eps} - \sigma_{+-}}{2 \eps} \right)^2 - \frac{2
    |\sigma_{+0}|^2}{\eps}} \hspace{0.5em} \le \hspace{0.5em}
  \sigma_{00}  \eqcm \nonumber \\ 
&& \sigma_{00} \hspace{0.5em} \le \hspace{0.5em}
\frac{\sigma_{\eps} - \sigma_{+-}}{2 \eps} + \sqrt{
  \left(\frac{\sigma_{\eps} - \sigma_{+-}}{2 \eps} \right)^2 - \frac{2
      |\sigma_{+0}|^2}{\eps}}  \eqpt
  \label{constraint2}
\end{eqnarray}
Note that the constraints in \eqref{constraint2} are the better the
larger $|\sigma_{+0}|$ is compared with $\sigma_{\eps} - \sigma_{+-}$.
This is what one would intuitively expect: a large interference term
between longitudinal and transverse polarisations implies that neither
$\sigma_{00}$ nor $\sigma_{++}$ can be very small, i.e.\ $\sigma_{00}$
can be neither very small nor very large compared with
$\sigma_{\eps}$. For $ |\sigma_{+0}| \ll \sigma_{\eps} - \sigma_{+-}$
the bounds \eqref{constraint2} are less stringent; they then differ
only by terms of relative order $|\sigma_{+0}|^2 / (\sigma_{\eps} -
\sigma_{+-})^2$ from the weaker (but simpler) conditions
\begin{eqnarray}
\frac{2 |\sigma_{+0}|^2}{\sigma_{\eps} - \sigma_{+-}} \le
\sigma_{00} \le \frac{\sigma_{\eps} - \sigma_{+-}}{\eps} \eqpt
  \label{constraint3}
\end{eqnarray}

By taking the derivative of \eqref{constraint1} to \eqref{constraint3}
with respect to $\eps$ one can see that all bounds are decreasing with
$\eps$ if the cross sections $\sigma_{++}$, $\sigma_{00}$ and
interference terms $\sigma_{+-}$, $\sigma_{+0}$ are kept fixed, so
that the lower bound is better for smaller $\eps$, i.e.\ larger $y$,
whereas the opposite holds for the upper bounds. Notice however that
at fixed $s$ a change in $y = (W^2 + Q^2 - m_p^2) /(s - m_p^2)$ means
a change in $W^2 + Q^2$ and will also change the $\sigma_{m n}$. If
their dependence on $W^2 + Q^2$ is only through a common global factor
then this factor drops out in the ratios between the bounds on
$\sigma_{00}$ and $\sigma_{00}$ itself.

Bounds of the form \eqref{constraint1} to \eqref{constraint3} can also
be derived for differential $\gp$ cross sections and interference
terms if they satisfy parity constraints analogous to
\eqref{ParityXsection}, i.e.\ if they depend only on parity invariant
variables. Thus one can obtain bounds on $\sigma_{00}$ by evaluating
inequalities analogous to \eqref{constraint1} and \eqref{constraint2}
for differential cross sections and then integrating them.

The usefulness of the bounds derived here depends of course on how
large the interference terms are. They will in general be better
in some parts of phase space than in others, a point we will
illustrate in section \ref{sec:resultsJ}.

An important point is that this method allows to {\em constrain\/} the
longitudinal cross section for {\em fixed\/} $\eps$, i.e.\ fixed $y$.
As is well known, a {\em measurement\/} of the longitudinal cross
section requires a variation of $y$, which means that one must either
measure the $ep$ cross section at different c.m.\ energies $\sqrt{s}$,
or, if $s$ is kept fixed, have information on how the transverse and
longitudinal $\gp$ cross sections depend on $W^2 + Q^2$.

We finally remark that up to now we have not used the requirement of a
fast outgoing proton or a rapidity gap between $X$ and the proton
remnant $\tilde{p}$ in reaction \eqref{DiffReac}. The analysis
developed here, and in particular the possibility to constrain the
longitudinal cross section, is directly applicable to ordinary deep
inelastic scattering. In fact, there has been much work on azimuthal
correlations in exclusive or semi-inclusive hadron production and in
semi-inclusive jet production \cite{OrdinaryDIS}, with the vector
$\vp$ defining the azimuthal angle $\varphi$ taken as the momentum of
the hadron or jet, resp. The $\varphi$-dependence of the $ep$ cross
section is always given by \eqref{XsectionMaster},
\eqref{MasterSimpler} with $\gp$ cross sections and interference terms
$\sigma_{m n}$ appropriately defined for the the process and angle
under consideration.

\section{Azimuthal dependence of dijet production}
\label{sec:jets}
\setcounter{equation}{0}

\subsection{Kinematics}
\label{sec:JetKin}

Our first example of an azimuthal angle in diffraction concerns events
where the proton is scattered elastically and the diffractive final
state $X$ consists of a quark-antiquark pair at parton level,
\begin{equation}
  e(k) + p(p) \to e(k') + p(\tilde{p}) + q(P_q) +
 \bar{q}(P_{\bar{q}})
 \eqcm
  \label{DiffJets}
\end{equation}
which hadronises into two jets. We allow for a finite mass $\mq$ of
the quark and antiquark. For the vector defining a direction in the
dijet system we choose
\begin{equation}
\vp = \qq = \frac{1}{2} (P_q - P_{\bar q})
  \label{QuarkDirection}
\end{equation}
and work in a reference frame where the incoming $p$ is collinear with
the $\gamma^\ast$, the photon momentum defining the $z$ axis, and
where the total momentum of the $q \bar{q}$-pair along this axis is
zero. It is related to the $\gp$ c.m.\ by a boost in $z$ direction, so
that azimuthal angles are the same in both frames. We introduce the
azimuthal angle $\phiqq$ between the electron momentum ${\bf k}$ and
${\bf \qq}$ as in sec.~\ref{sec:single}, and the azimuthal angle $\dq$
between ${\bf \qq}$ and ${\bf \Delta}$, where $\Delta = p - \tilde{p}$
is the momentum transfer from the proton. We will integrate over $\dq$
in the present section. Another useful variable is the longitudinal
component $\Pl$ of ${\bf \qq}$, and thus of ${\bf P}_q$, along the
photon direction. Its range is from $- \Pl^{\it max}$ to $\Pl^{\it
  max}$ with $\Pl^{\it max} = \sqrt{\Mx^2 / 4 - \mq^2}$, and thus
independent of $\dq$. This is in contrast to the length of the
transverse part $\Pt$ of ${\bf \qq}$ since
\begin{equation}
\Pt^2 = \frac{\Mx^2 / 4 - \mq^2 - \Pl^2}{1 + t /(\Mx^2 - t) \cdot
  \cos^2 \dq}  \eqpt
  \label{QQtransverse}
\end{equation}
Only for $t = 0$ do the transverse momenta of $q$ and $\bar{q}$
balance; then $\Pt^2$ is just the squared transverse jet momentum.
$\Pl$ is parity invariant, which will allow us to use the simplified
expression \eqref{MasterSimpler} instead of \eqref{XsectionMaster} to
obtain the $\phiqq\,$-dependence of the $ep$ cross section.

Experimentally it is difficult to establish which of the two jets
originated in the quark $q$ and which in the antiquark $\bar{q}$. It
is therefore useful to sum over final states where the momenta of $q$
and $\bar{q}$ are interchanged, in other words over $\qq$ and $-\qq$.
One can show that after this symmetrisation the transverse and
longitudinal $\gp$ cross sections and the transverse-transverse
interference terms are even in $\Pl$, whereas the
transverse-longitudinal interference terms are odd in $\Pl$ and vanish
at $\Pl = 0$.

To define an azimuthal angle after summing over $\qq$ and $-\qq$ one
can distinguish the two jets kinematically, e.g.\ according to which
one points in the forward direction with respect to the photon. Let
$P_F$ be the four-momentum of the forward and $P_B$ that of the
backward jet, and choose for the direction $\vp$ instead of
\eqref{QuarkDirection}
\begin{equation}
\vp = \qj = \frac{1}{2} (P_F - P_B) 
  \label{JetDirection}
\end{equation}
with the corresponding relative azimuthal angle $\phifb$ between ${\bf
  k}$ and ${\bf P}_{\it FB}$. The longitudinal component of ${\bf
  P}_{\it FB}$ is $\Pfb$. Writing $\qj = {\rm sgn}(\Pl) \cdot \qq$ we
see that $\qj$ is a polar vector so that we can again use
\eqref{MasterSimpler} for the $ep$ cross section.

In the case $\Pl = 0$, i.e.\ when the jet momenta are perpendicular to
the $\gp$ axis, eq.~\eqref{JetDirection} leaves the sign of $\vp$
undefined, as the attribution of $P_F$ and $P_B$ to the jets is
ambiguous. This leads to an ambiguity between azimuthal angles
$\phifb$ and $\phifb + \pi$ and hence a to sign ambiguity for
$\cos\phifb$ and $\sin\phifb$ but not for $\cos(2 \phifb)$ and $\sin(2
\phifb)$. As mentioned above the transverse-longitudinal interference
terms vanish at $\Pl = 0$ when summed over $\qq$ and $-\qq$, so that
no ambiguity remains in the $ep$ cross section.

The summation over $\qq$ and $-\qq$ is in fact trivial under the
assumption that the diffractive exchange has definite charge
conjugation parity, which is of course the case for pomeron or
two-gluon exchange. Applying charge conjugation invariance of the
strong interactions to the photon dissociation part of the $\gp$
subreaction it then follows that the differential cross section for
\eqref{DiffJets} remains the same if we exchange $q$ and $\bar{q}\,$:
\begin{equation}
d \sigma\left(e p \to e p + q(P_q) + \bar{q}(P_{\bar{q}})\right) = 
d \sigma\left(e p \to e p + \bar{q}(P_q) + q(P_{\bar{q}})\right) \eqpt
  \label{exchange}
\end{equation}
To sum over $\qq$ and $-\qq$ and change from the variable $\qq$ to
$\qj$ we thus only need to multiply the cross section with 2 and
replace $\phiqq$ with $\phifb$ and $\Pl$ with $\Pfb$.

\subsection{The dijet cross section at $t = 0$ in the
  Landshoff-Nachtmann model}
\label{sec:zero}

In this subsection we give the differential cross section for reaction
\eqref{DiffJets} at $t = 0$ in the model of Landshoff and Nachtmann
(LN) \cite{LN,DL}. The transverse and longitudinal $\gp$ cross
sections have been computed in \cite{MD}, and the calculation of the
interference terms goes along the same lines. We therefore only recall
the essentials of the model and give the final results.

The LN model was developed to give a simple QCD based description of
the soft pomeron. It approximates pomeron exchange by the exchange of
two gluons which are taken as nonperturbative, i.e.\ they have a
nonperturbative propagator $-g_{\mu \nu} D(l^2)$ instead of $-g_{\mu
  \nu} / l^2$ in Feynman gauge. The nonperturbative propagator
$D(l^2)$ is a difficult quantity to compute and indeed there is no
consensus in the literature about its behaviour at small $l^2$
\cite{Propagator}, but we will not attempt to discuss this issue here.
We will instead follow the rather model independent approach of
\cite{LN,DL,Alpha}, which is based on the observation that often the
amplitude of a considered process can be approximated in such a way
that it depends on $D(l^2)$ only via certain simple integrals, so that
it is not necessary to know the detailed functional form of $D(l^2)$.
In the present paper we will need the two moments
\begin{eqnarray}
  \int_{0}^{\infty} d l^{2} [\alpha_{s}^{(0)} D(-l^{2})]^{2} &=&  
  \frac{9 \beta_{0}^{2}}{4\pi} \eqcm              \label{moment1}  \\
  \int_{0}^{\infty} d l^{2} [\alpha_{s}^{(0)} D(-l^{2})]^{2} 
  \cdot l^{2} &=&  
  \frac{9 \beta_{0}^{2} \mu_{0}^{2}}{8\pi} \eqcm  \label{moment2}
\end{eqnarray}
where $\beta_{0} \approx 2.0 \, {\rm GeV}^{-1}$ and $\mu_{0} \approx
1.1 \, {\rm GeV}$ have been estimated from data \cite{DL,Alpha}.  From
the ratio of \eqref{moment2} and \eqref{moment1} $\mu_0^2$ appears as
the scale characteristic for the behaviour of $D$ on $l^2$. Following
\cite{Alpha} we take $\alpha_{s}^{(0)} \approx 1$ for the strong
coupling in the nonperturbative region which dominates the
$l^2$-integrations in \eqref{moment1}, \eqref{moment2}.

In this model the reaction $\gp \to q \bar{q} \, p$ is described by
the exchange of two gluons between a quark in the proton and the $q
\bar{q}$-pair into which the virtual photon splits. In the high-energy
limit its amplitude is purely imaginary and thus can be calculated by
cutting the corresponding Feynman diagrams in the $s$-channel. In each
diagram there is then exactly one off-shell quark, namely one of the
quarks into which the photon dissociates. Its typical virtuality is
\begin{equation}
\lambda^2 = \frac{\Pt^2 + \mq^2}{1 - \beta} = 
            (\Mx^2 + Q^2) \frac{\Pt^2 + \mq^2}{\Mx^2} \eqcm
  \label{BartelsScale}
\end{equation}
which can be seen as the relevant scale of hardness of the process
\cite{BarLottWust,NikolaevCharm}. We obtain the $ep$ cross section
from the master equation \eqref{MasterSimpler} as
\begin{eqnarray}
\lefteqn{ \frac{d \sigma(e p \to e p \, q \bar{q})}{d \phiqq \, d x\,
  d Q^2 \, d \Pl \, d \beta \, dt} = \frac{\alpha_{\it em}}{\pi} \, 
  \frac{1}{2 \pi x \, Q^2} \left( 1 - y + y^2 /2 \right) 
\left\{\frac{d \sigma^{q\bar{q}}_{++}}{d \Pl \, d \beta \, d t} +
  \eps \frac{d \sigma^{q\bar{q}}_{00}}{d \Pl \, d \beta \, d t}
  \right. } \nonumber \\  
& &  \left. -  \eps\cos(2\phiqq)\ \frac{d \sigma^{q\bar{q}}_{+-}}{d
  \Pl \, d \beta \, d t} - 2 \sqrt{\eps(1+\eps)}\cos\phiqq \ 
  \frac{d \sigma^{q\bar{q}}_{+0}}{d \Pl \, d \beta \, d t} \right\}
  \eqpt \hspace{6em}
  \label{SigmaJetsElectron}
\end{eqnarray}
$d \sigma^{q\bar{q}}_{+0} / (d \Pl \, d \beta \, d t)$ is purely real
in our approximation since the $\gp$ amplitude is purely imaginary, so
that its phase is independent of the photon polarisations, see our
discussion in sec.~\ref{sec:single}. For the differential $\gp$ cross
sections and interference terms we find
\begin{equation}
  \left. \frac{d \sigma^{q\bar{q}}_{mn}}{d \Pl \, d \beta \, d t}
  \right|_{t = 0} 
  =  \frac{8}{3} \, \alpha_{\it em} e_q^2 \,  
  \frac{\alpha_{s}(\lambda^2)}{\alpha_{s}^{(0)}} \,
  \xi^{2 (1 - \alpha_\pom(0))} \, \frac{1 - \beta}{\beta \Mx (\Mx^2 +
  Q^2)^2}
  \, {\cal S}^{q\bar{q}}_{m n}  \eqcm
  \label{SigmaJetsGamma}
\end{equation}
where $e_q$ is the electric charge of the produced quark in units of
the positron charge and $\alpha_{\pom}(t) = 1 + \epsilon + \alpha' t$
with $\epsilon \approx 0.085$ and $\alpha' \approx 0.25 \GeV^{-2}$ the
soft pomeron trajectory. The reduced cross sections
\begin{eqnarray}
{\cal S}^{q\bar{q}}_{++} &=&
\left(1 - 2\, \frac{\Pt^2 + \mq^{2}}{\Mx^2} \right) \,
  \frac{\Pt^2}{\Pt^2 + \mq^{2}} \, (\Mx^2 + Q^2)^2 \, L_1(\Pt^2,
  \iw)^2 \nonumber \\
 & & + \frac{\mq^{2}}{\Pt^2 + \mq^{2}} \, (\Mx^2 + Q^2)^2 \,
 L_2(\Pt^2, \iw)^2  \eqcm \nonumber \\  
{\cal S}^{q\bar{q}}_{00} &=& 4 \, \frac{Q^{2}}{\Mx^2} \, \frac{\Pt^2 +
    \mq^{2}}{\Mx^2} \, (\Mx^2 + Q^2)^2 \, L_2(\Pt^2, \iw)^2  \eqcm
  \nonumber \\ 
{\cal S}^{q\bar{q}}_{+-} &=& 2 \, \frac{\Pt^2 + \mq^{2}}{\Mx^2}
\frac{\Pt^2}{\Pt^2 
  + \mq^{2}} \, (\Mx^2 + Q^2)^2 \, L_1(\Pt^2, \iw)^2 
  \eqcm \nonumber \\ 
{\cal S}^{q\bar{q}}_{+0} &=& - 2 \sqrt{2} \, \frac{Q}{\Mx} \,
\frac{\Pl | \Pt |}{\Mx^2} \, (\Mx^2 + Q^2)^2 \, L_1(\Pt^2, \iw)
L_2(\Pt^2, \iw)  \eqcm
  \label{Imn} 
\end{eqnarray}
whose normalisation  has been chosen for later convenience, involve
loop integrals
\begin{equation}
L_i(\Pt^2, \iw) = \int_{0}^{\infty} d \lt^2 \,
 [\alpha_{s}^{(0)} D(- \lt^2)]^{2} \, f_{i}(\iv, \iw) \eqcm 
  \hspace{4em}  i = 1,2
  \label{LoopIntegrals}
\end{equation}
over the functions 
\begin{eqnarray}
f_1(\iv, \iw) &=& 1 - \frac{1}{2 \iw} \left[ 1 - \frac{\iv + 1 - 2
  \iw}{\sqrt{(\iv + 1 - 2 \iw)^2 + 4 \iw (1 - \iw)}} \right]  \eqcm
  \nonumber \\ 
f_2(\iv, \iw) &=& 1 - \frac{1}{\sqrt{(\iv + 1 - 2 \iw)^2 + 4 \iw (1 -
  \iw)}} 
  \label{functions}
\end{eqnarray}
of the dimensionless variables\footnote{The definitions of $\iv$ and
  $\iw$ here differ from those in \protect\cite{MD}.}
\begin{equation}
\iv = \frac{\lt^2}{\lambda^2}  \eqcm \hspace{5em}
\iw = \frac{\Pt^2}{\lambda^2} = (1 - \beta) \, \frac{\Pt^2}{\Pt^2 +
  \mq^2} \eqpt
  \label{variables}
\end{equation}

Assuming that due to the squared gluon propagator the dominant values
of $\lt^2$ in the loop integrals $L_i$ are small compared with
$\lambda^2$ we can Taylor expand $f_i(\iv, \iw)$ at $\iv = 0$ and
approximate
\begin{equation}
f_i(\iv, \iw) \approx \iv \cdot \left. \frac{\partial f_i(\iv,
  \iw)}{\partial  \iv} \right|_{\iv = 0} \eqcm \nonumber \\ 
  \label{LoopApprox}
\end{equation}
so that the remaining integral is given by \eqref{moment2}. The
integral \eqref{moment1} does not appear because both $f_1$ and $f_2$
vanish at $\iv = 0$. In this approximation \eqref{Imn} becomes
\begin{eqnarray}
{\cal S}^{q\bar{q}}_{++} &=& \left( \frac{9 \beta_0^2 \mu_0^2}{8 \pi}
\right)^2 
 \left\{ 4 \, \left( \frac{\Mx^2}{\Pt^2 + \mq^2}
 \right)^2  \left( 1 - 2\, \frac{\Pt^2 + \mq^2}{\Mx^2} \right)
  \frac{\Pt^2}{\Pt^2 + \mq^2} \, (1 - \iw)^2 \right. \nonumber \\
 & &  \phantom{\left( \frac{9 \beta_0^2 \mu_0^2}{8 \pi} \right)^2}
   + \left. \left( \frac{\Mx^2}{\Pt^2 + \mq^2} \right)^2 
   \frac{\mq^2}{\Pt^2 + \mq^2} \, (1 - 2 \iw)^2 \right\}  \eqcm
 \nonumber \\
{\cal S}^{q\bar{q}}_{00} &=& \left( \frac{9 \beta_0^2 \mu_0^2}{8 \pi}
 \right)^2 \cdot 4 \, \frac{Q^2}{\Mx^2} \cdot \frac{\Mx^2}{\Pt^2 +
 \mq^2} \, (1 - 2 \iw)^2   \eqcm \nonumber \\
{\cal S}^{q\bar{q}}_{+-} &=&  \left( \frac{9 \beta_0^2 \mu_0^2}{8 \pi}
 \right)^2 
 \cdot 8 \, \frac{\Mx^2}{\Pt^2 + \mq^2} \cdot
  \frac{\Pt^2}{\Pt^2 + \mq^2} \,  (1 - \iw)^2  \eqcm \nonumber \\
{\cal S}^{q\bar{q}}_{+0} &=& - \left( \frac{9 \beta_0^2 \mu_0^2}{8
 \pi} \right)^2 
 \cdot 4 \sqrt{2} \, \frac{Q}{\Mx} \cdot
 \frac{\Mx^2}{\Pt^2 + \mq^2} \, \frac{\Pl |\Pt|}{\Pt^2 + \mq^2} \, 
  (1 - \iw) \, (1 - 2 \iw)  \eqpt \nonumber \\
  \label{Imn2}
\end{eqnarray}

As a benchmark we have compared the integrals $L_i(\Pt^2, \iw)$ in the
approximation \eqref{LoopApprox} with their exact values for the model
gluon propagator used in \cite{DL}:
\begin{equation}
  D(- l^2) \propto \left[1+\frac{l^2}{(n-1) \mu_{0}^{2}}
  \right]^{-n} \eqcm \hspace{4em} n \ge 4 \eqcm
  \label{SpecialGluon}
\end{equation}
where the proportionality constant can easily be obtained from
\eqref{moment1}. For $n \rightarrow \infty$ this becomes $D(- l^2)
\propto \exp(- l^2 / \mu_{0}^{2})$. We find that the value of $n$ has
little influence on the $L_i(\Pt^2, \iw)$, and that the approximations
\eqref{LoopApprox} are in general rather good, except however for some
regions of parameter space. In particular the approximation of $L_1$
becomes bad for $\iw$ close to 1 and for small $\Pt^2$.  On the other
hand $L_2$ becomes zero and changes its sign for some $\Pt^2$ if $\iw
> 1 /2$ because the function $f_2(\iv)$ changes sign at $\iv = \iw
\cdot \lt^2 / \Pt^2 = 2 (2 \iw - 1)$. The parameters $\Pt^2, \iw$ for
which $L_2$ vanishes are not always well reproduced by the
approximation; it can be seen that for given $\Pt^2$ the value $\iw
=1/2 $ obtained from \eqref{LoopApprox} is too small, so that with
fixed $\mq^2$ the corresponding value of $\beta$ is overestimated

An improved approximation, also leading to the moment \eqref{moment2},
is achieved by replacing \eqref{LoopApprox} with
\begin{equation}
f_i(\iv, \iw) \approx \iv \cdot \frac{f_i(\iv_0, \iw)}{\iv_0}
\eqcm \hspace{4em} \iv_0 = \frac{l_0^2}{\lambda^2} \eqcm 
  \label{ApproxImprove}
\end{equation}
where we take $\l_0^2 = \mu_0^2$. With this approximation the values
$\Pt^2, \iw$ where $L_2$ vanishes are reproduced much better, and the
errors on $L_1$ are in the region of a few percent even if $\iw = 0.9$
and $\Pt^2$ as small as $2 {\rm \ GeV}^2$.

\subsection{Discussion of the results}
\label{sec:resultsJ}

Let us make some remarks on the results \eqref{Imn}, \eqref{Imn2}.
The first concerns the sign of the $\gp$ interference terms. The
transverse-transverse interference is always positive, so that the
term with $\cos(2 \phiqq)$ in the $ep$ cross section
\eqref{SigmaJetsElectron} is negative. In \cite{BarLott} it was
pointed out that this is the opposite sign than the one obtained for
$q \bar{q}$-production in photon-gluon fusion. The sign of the
longitudinal-transverse interference depends on the loop integral
$L_2(\Pt^2, \iw)$ and thus on the value of $\iw$. \eqref{Imn2} gives a
sign change at $\iw = 1 /2$, the exact value of $\iw$ from \eqref{Imn}
is larger and depends on $\Pt^2$ as mentioned at the end of the
previous subsection. This characteristic change of sign has also been
observed in \cite{BarLott}. As a general remark we can say that the
distribution of the $e p$ cross section in $\phiqq$ we obtain is very
similar to the one in the perturbative two-gluon approach of
\cite{BarLott}, apart from the overall normalisation which comes out
different in the two models, cf.~\cite{BarEtAl}. The main
characteristics of the normalised azimuthal distribution are
determined by the two-gluon exchange picture.

We now turn to the dependence on the transverse jet momentum. From
\eqref{Imn} one sees that compared with the transverse cross section
the transverse-longitudinal interference is suppressed by a factor
$|\Pt| / \Mx$, the transverse-transverse interference by $\Pt^2 /
\Mx^2$ and the longitudinal cross section by \linebreak $(\Pt^2 +
\mq^2)/ \Mx^2$. This means that the interferences (and the
longitudinal cross section) are less important if $\Pt^2$ is small
compared with $\Mx^2$, i.e.\ if the jets are close to the $\gp$ axis
in the reference frame we are working in. Note that for light quarks
the suppression of the transverse-longitudinal interference is weaker
than the one for the longitudinal cross section. This suggests a way
to experimentally look for the zero in the longitudinal $\gp$
amplitude, which is due to the behaviour of the integral $L_2(\Pt^2,
\iw)$ and can be viewed as a characteristic feature of the two-gluon
exchange mechanism in this reaction: The zero might be seen through
the change of sign of the $\cos \phiqq\,$-term in the angular
dependence as $w$ or $\beta$ is varied, whereas it should be difficult
to observe it from a dip in the $\phiqq\,$-integrated spectra, given
that the longitudinal $\gp$ cross section is much smaller than the
transverse one where the zero occurs. We finally note that for heavy
quarks \eqref{Imn} is valid down to $\Pt^2 = 0$ and that all
interference terms (though not the longitudinal cross section) vanish
in this limit where $\vp$ is collinear with $p$ and $q$ as required by
angular momentum conservation.

\begin{figure}
    \begin{center}
    \leavevmode
    \epsfxsize=0.95\textwidth  \epsfbox[87 123 493 636]{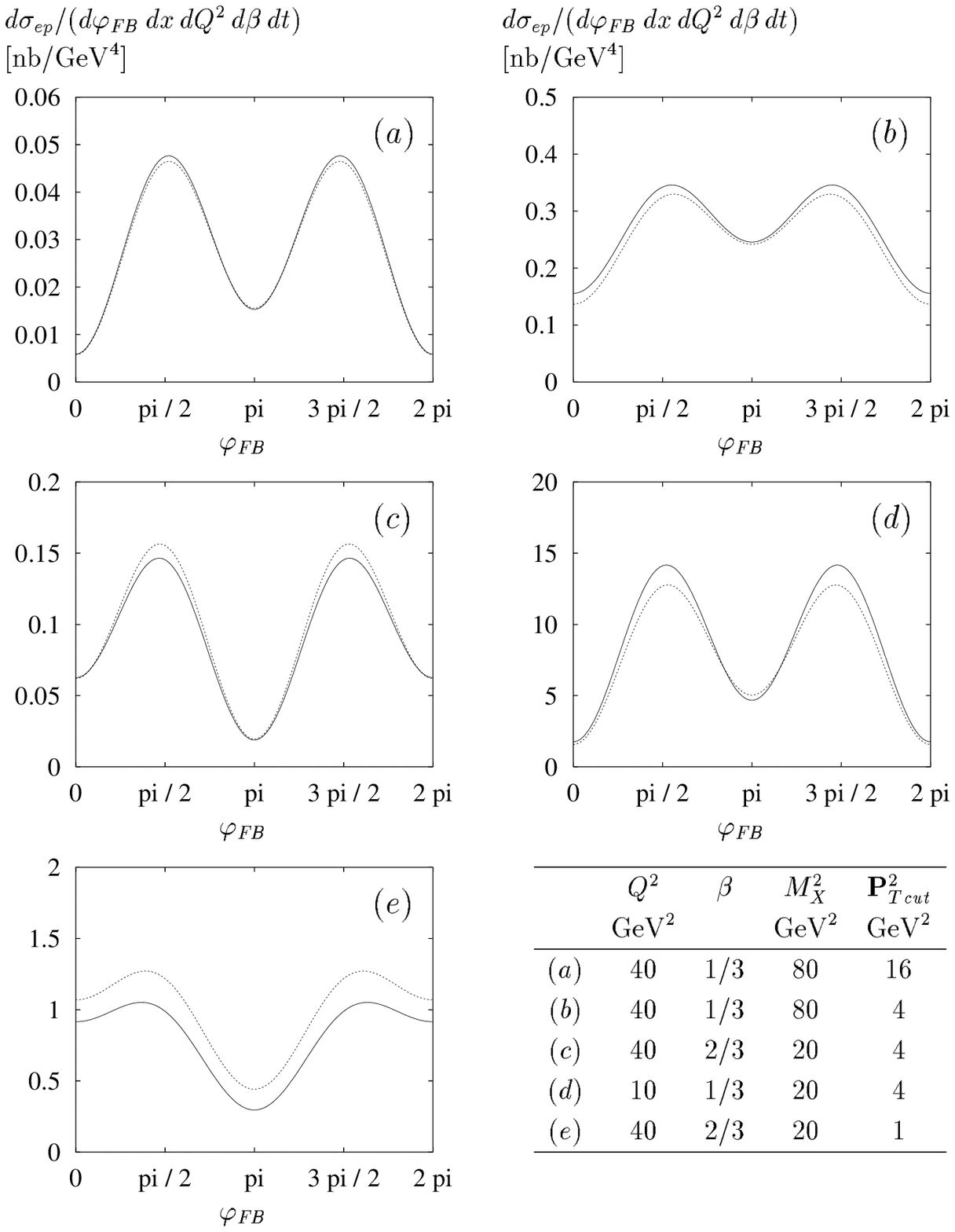}
  \end{center}
  \caption{\label{fig:AngleJet}Dependence on $\phifb$ of $d \sigma(ep
    \to ep \, q \bar{q}) / (d \phifb \, d x\, d Q^2 \, d \beta \,
    dt)$, summed for $u,d,s$ and $c$ quarks. Values of the free
    kinematical variables are $\protect\sqrt{s} = 296 {\rm \ GeV}$, $y
    = 0.5$, $\eps = 0.8$, $t = 0$ and those given in the table. A cut
    $\xi \le 0.05$ has been imposed. Full lines correspond to the
    improved approximation (\protect\ref{ApproxImprove}) of
    (\protect\ref{Imn}), dotted ones to the simplified results
    (\protect\ref{Imn2}) obtained with the approximation
    (\protect\ref{LoopApprox}).}
\end{figure}

For a numerical study we change variables from $\phiqq$ and $\Pl$ to
$\phifb$ and $\Pfb$ as explained at the end of sec.~\ref{sec:JetKin}
and integrate over $\Pfb$. To ensure that we have jets and that the
scale $\lambda^2$ of \eqref{BartelsScale} remains large we impose a
lower cut $\Ptcut^2$ on $\Pt^2$, which at $t = 0$ corresponds to an
$\mq^2$-dependent upper cut $\Plcut$ on $\Pfb$,
see~\eqref{QQtransverse}. The $\phifb$-dependence of the $ep$ cross
section for the quark flavours $u, d, s, c$ at the HERA c.m.\ energy
of $\sqrt s = 296 {\rm \ GeV}$ is shown in fig.~\ref{fig:AngleJet},
where we plot $d \sigma(ep \to ep \, q \bar{q}) / (d \phifb \, d x\, d
Q^2 \, d \beta \, dt)$ as a function of $\phifb$ for different values
of the other kinematical variables. We note that in the examples where
charmed jets can be produced kinematically their fraction in the $\gp$
cross section is not negligible; for cases $(a)$ and $(b)$ of the
table in fig.~\ref{fig:AngleJet} it is about $1/3$, and for case $(e)$
about $1/5$.


The examples in fig.~\ref{fig:AngleJet} illustrate how a smaller
minimum $\Pt^2 / \Mx^2$ leads to a flatter dependence on $\phifb$ as
discussed above, while increasing the overall rate. The effect of
$\beta$ on the sign of the $\cos\phifb$-term in the cross section can
clearly be seen. Also shown in the plots is the difference between the
approximations \eqref{ApproxImprove} and \eqref{LoopApprox} of the
integrals in \eqref{Imn}: in general the less exact approximation
\eqref{LoopApprox} which leads to \eqref{Imn2} is rather good,
especially if $\Ptcut^2$ is large.

Finally we investigate what bounds on the longitudinal $\gp$ cross
section one could obtain from measuring the $\phifb$-dependence shown
in fig.~\ref{fig:AngleJet}. For convenience we introduce the
quantities
\begin{eqnarray}
F^{\it FB}_{++} &=& \frac{d \sigma^{\it FB}_{++}}{d \beta \, d t}
\eqcm \hspace{2em}
F^{\it FB}_{00} \arreq \eps \frac{d \sigma^{\it FB}_{00}}{d \beta \, d
  t} \eqcm \hspace{2em}
F^{\it FB}_{\eps} \arreq F^{\it FB}_{++} + F^{\it FB}_{00} 
\eqcm \nonumber \\
F^{\it FB}_{+-} &=& - \eps \frac{d \sigma^{\it FB}_{+-}}{d \beta \, d
  t} \eqcm \hspace{2em}
F^{\it FB}_{+0} \arreq -2 \sqrt{\eps (1+\eps)} \, \frac{d \sigma^{\it
    FB}_{+0}}{d \beta \, d t}  \eqcm
  \label{FourierJets}
\end{eqnarray}
whose factors are chosen such that
\begin{eqnarray}
\lefteqn{ \frac{d \sigma(e p \to e p \, q \bar{q})}{d \phifb \, d x\,
  d Q^2 \, d \beta \, dt} = \frac{\alpha_{\it em}}{\pi} \, 
  \frac{1}{2 \pi x \, Q^2} \left( 1 - y + y^2 /2 \right) } 
  \nonumber \\
&& \left\{ F^{\it FB}_{\eps} + F^{\it FB}_{+-} \cos(2\phifb) + 
  F^{\it FB}_{+0} \cos\phifb \right\} 
  \eqpt 
  \label{FourierInSigma}
\end{eqnarray}
Up to a global factor $F^{\it FB}_{\eps}$, $F^{\it FB}_{+0}$ and
$F^{\it FB}_{+-}$ are therefore the Fourier coefficients for the
$\phifb$-dependence of the $ep$ cross section, and $F^{\it FB}_{00}$
is the contribution of longitudinal photons to $F^{\it FB}_{\eps}$.
Table \ref{tab:JetBounds050} gives the coefficients $F^{\it FB}_{+-}$,
$F^{\it FB}_{+0}$ and $F^{\it FB}_{\eps}$ which correspond to the
plots in fig.~\ref{fig:AngleJet}, the longitudinal contributions
$F^{\it FB}_{00}$, and the lower and upper bounds $F^{\it FB}_{\it
  low}$, $F^{\it FB}_{\it upp}$ on $F^{\it FB}_{00}$ one can obtain
from the $\phifb$-dependence using the differential analogues of
\eqref{constraint2}. The upper bound \eqref{constraint1} is not useful
since the transverse-transverse interference $d \sigma^{\it FB}_{+-}
/(d \beta\, d t)$ is positive.

When the minimum $\Pt^2 / \Mx^2$ is rather small, i.e.\ in cases $(b)$
and $(e)$, the lower and upper bounds are rather far apart from each
other and in this sense not very stringent, due to the suppression of
the transverse-longitudinal interference by $|\Pt| / \Mx$ compared
with the transverse cross section. From \eqref{constraint3} we see
that the lower bound is then suppressed by $\Pt^2 / \Mx^2$. On the
other hand the longitudinal cross section itself has a suppression
factor $(\Pt^2 + \mq^2)/ \Mx^2$, and as a result the lower bound we
obtain is quite close to the actual value of $F^{\it FB}_{00}$.

\renewcommand{\arraystretch}{1.2}
\begin{table}
  \caption{\label{tab:JetBounds050}Fourier coefficients corresponding
    to the angular distributions shown in
    fig.~\protect\ref{fig:AngleJet} and lower and upper bounds  
    $F^{\it FB}_{\it low}$, $F^{\it FB}_{\it upp}$ on the
    longitudinal contribution $F^{\it FB}_{00}$ one can
    obtain from them. For the definition of $F^{\it FB}_{\eps}$,
    $F^{\it FB}_{00}$ etc.\ cf.~(\protect\ref{FourierJets}).  $y =
    0.5$, $\eps = 0.8$, $\protect\sqrt{s} = 296 {\rm \ GeV}$, $t =
    0$.}
  \begin{center}
    \leavevmode
    \begin{tabular}{cccccccccc}  \hline
      & $\beta$ & $\Mx^2$ & $\Ptcut^2$ & $F^{\it FB}_{+-}$ & 
        $F^{\it FB}_{+0}$ & $F^{\it FB}_{\eps}$ & $F^{\it FB}_{00}$ &
        $F^{\it FB}_{\it low}$ & $F^{\it FB}_{\it upp}$ \\
      & & ${\rm GeV}^2$ & ${\rm GeV}^2$ & \multicolumn{6}{c}{\dotfill
        ${\rm \ nb} / {\rm GeV}^2$} \dotfill \\  \hline
$(a)$ & 1/3 & 80 & 16 & -2.9 & -0.75 &  4.6 & 0.45 & 0.22 & 0.72 \\
$(b)$ & 1/3 & 80 & 4  &  -11 &  -7.1 &   43 &  1.3 & 0.50 &   28 \\
$(c)$ & 2/3 & 20 & 4  & -8.3 &   3.5 &   15 &  2.6 &  1.0 &  3.3 \\
$(d)$ & 1/3 & 20 & 4  &  -54 &   -14 &   86 &  7.2 &  4.1 &   14 \\
$(e)$ & 2/3 & 20 & 1  &  -30 &    49 &  130 &   18 &  8.4 &   80 \\
\hline
    \end{tabular}
  \end{center}
\end{table}

\renewcommand{\arraystretch}{1}

\subsection{Jet angle for more general final states}
\label{sec:general}

We now generalise the jet angle used so far to diffractive final
states $X$ that do not necessarily have a two-jet topology. As before
we work in a reference frame where the incoming photon and proton are
collinear and where the total momentum of $X$ along this axis is zero.
Let {\boldmath $\tau$} be the thrust axis of $X$ in this frame. It
can be oriented by requiring that it points in the direction of the
photon momentum:
\begin{equation}
\mbox{\boldmath $\tau$} = \mbox{\boldmath $\tau$} \, 
{\rm sgn}({\bf q} \pdot \mbox{\boldmath $\tau$})  \eqpt
  \label{OrientThrust}
\end{equation}
This provides a direction in the hadronic final state, which we can
also write as a four-vector:
\begin{equation}
\vp = (0, \mbox{\boldmath $\tau$})   \eqpt
  \label{CovariantThrust}
\end{equation}
From \eqref{OrientThrust} and \eqref{CovariantThrust} it follows that
$\vp$ is a polar vector. Note that in the case of a two-jet final
state and in the limit $t = 0$ it becomes proportional to the vector
$\qj$ defined in \eqref{JetDirection}. Another possibility would be to
define $\vp$ from the thrust axis in the rest frame of $X$ by
equations analogous to \eqref{OrientThrust} and
\eqref{CovariantThrust}, and then to boost $\vp$ to the $\gp$ frame.
If $X$ is a dijet this is then proportional to $\qj$ even at finite
$t$.

The vector $\vp$ defined in one of these ways, or a vector obtained
from another suitable shape variable of the system $X$, can be used
for the definition of the azimuthal angle $\varphi$ and of $\gp$ cross
sections and interference terms. It is not restricted to events with
only two jets in $X$, and it does not require to have jets with a
transverse momentum large enough for a jet algorithm to be applicable.
This could allow for a significant gain in statistics. The measurement
of the $\varphi$-dependence could in particular be used to constrain
the cross section for longitudinal photons. The discussion in the
previous subsection and the numerical example with $\Pt^2 \ge 1 {\rm \ 
  GeV}^2$ in table~\ref{tab:JetBounds050} indicates that one might
obtain at least a useful lower bound even for low $\Pt^2$, provided
the ratio $\Pt^2 / \Mx^2$ is not very small. Too small values of
$\Pt^2 / \Mx^2$ will presumably also present experimental problems,
since then the polar angle of $\vpb$ is close to zero and the
resolution on its azimuth will become poor.

When the thrust axis is perpendicular to the $\gp$ direction the
requirement \eqref{OrientThrust} does not fix its orientation, so that
the angle $\varphi$ is only defined up to an ambiguity between
$\varphi$ and $\varphi + \pi$. This is just as in the case $\Pl = 0$
for the two-jet final state which was discussed at the end of
sec.~\ref{sec:JetKin}. Using a similar argument as there one can show
that the transverse-longitudinal interference terms vanish when the
thrust axis is perpendicular to ${\bf q}$ so that no ambiguity appears
in the $ep$ cross section.

\section{Dijet production at finite $t$}
\label{sec:finite}
\setcounter{equation}{0}

\subsection{Coupling of the two gluons to the proton}
\label{sec:lower}

We will now investigate diffractive production of a jet pair
\eqref{DiffJets} at finite $t$ in the LN model. Throughout our
calculation we take the high energy limit, dropping terms that are
suppressed by factors of $\xi$. In this approximation $t = \Delta_T^2$
where $\Delta_T$ is the transverse part of $\Delta$ with respect to
$p$ and $q$. A characteristic property of the LN model is that the two
gluons couple to the same quark in a hadron \cite{LN}. The coupling of
the gluons to the proton is then given by the isoscalar vector current
of the nucleon, and the squared amplitude for the process is
proportional to \cite{HDCB}
\begin{equation}
\tilde{G}^2(t) = F_1^2(t) - \frac{t}{4 m_p^2} F_2^2(t)  \eqpt
  \label{ProtonFormf}
\end{equation}
where $F_1(t)$ and $F_2(t)$ are the isoscalar Dirac and Pauli form
factors of the nucleon, respectively, i.e.\ the sum of the Dirac
(Pauli) form factors of the proton and the neutron. At $t = 0$ one has
$F_1(0) = 1$ and $F_2(0) \approx - 0.12$, cf.~\cite{HDCB}, and in the
region $|t| \lsim 1 {\rm \ GeV}^2$ we are interested in $F_1(t)$ is
dominating this expression.

One finds that at high energy the relevant kinematics in the Feynman
diagrams for $\gp \to q \bar{q} \, p$ are determined by the proton
momentum and the kinematic variables of the $\gamma^\ast \to q
\bar{q}$ transition, but not on the momentum of the quark within the
proton. All dependence of the amplitude on the nucleon structure thus
comes from the form factor $\tilde{G}(t) \,$; there is no further
dependence on tranverse or longitudinal momentum distributions of
quarks, even at finite $\Delta_T$.

The polarisation vectors for the gluons coupling to the proton both
come out proportional to the initial proton momentum $p$. For the
complete amplitude they are contracted with the propagators of the
gluons, and the result is contracted with a tensor corresponding to
the gluons coupling to the produced quark and antiquark. We wish to
remark that one need not take a Feynman-like gauge for the gluons,
i.e.\ a propagator $- g_{\mu \nu} D(k^2)$ where $k$ is the gluon
momentum. In fact one has some freedom to choose a gauge in our model
without changing the structure of the result: there is no contribution
to the amplitude from the tensor $k_\mu k_\nu$ in a general covariant
gauge, nor from $k_\mu n_\nu + n_\mu k_\nu$ with some fixed
four-vector $n$, which appears in non-covariant gauges. The reason is
that in the approximation of our calculation the exchanged gluons
couple directly to quarks, not to gluons, and thus to a conserved
vector current.  Note however that for terms in the propagator which
involve $n_\mu n_\nu$ (appearing in radiative corrections to the bare
propagator in non-covariant gauges) one would have to investigate in
detail whether the extra tensors contribute to the leading energy
behaviour of the amplitude.

The approximation of two noninteracting gluons in the LN model is
certainly a very crude one. To go beyond it one could replace the
direct coupling of the gluons to a quark in the proton by the cut
amplitude for the emission by the proton of two gluons, in other words
by the cut amplitude for $g^\ast p \to g^\ast p$. Including the gluon
propagators the latter might be approximated by the gluon distribution
in the proton \cite{GluonsInProton,BarLottWust} at zero $t$. This is
however not useful when we want to compute the effects of finite $t$,
since in the gluon distribution the four-momenta of the two gluons are
by definition equal, in particular they do not transfer any transverse
momentum $\Delta_T$.

Some features of the LN model are also found in this more general
framework if one makes the assumption that the squared c.m.\ energy of
the $g^\ast p \to g^\ast p$ amplitude is small compared to $W^2$ in
the region of phase space which dominates the amplitude for $\gp \to q
\bar{q} \, p$. This is for instance the case in the multiperipheral
approximation.  Then one can show that the polarisation of the gluons
is again proportional to $p$, and that the relevant kinematics in the
Feynman diagrams are as in the LN model calculation.  Moreover, both
statements remain valid if the proton dissociates and one integrates
over the particle momenta in the proton remnant with $\tilde{p}$ being
held fixed, provided that $\tilde{p}^2 \ll W^2$.

What is however particular to our model is the dependence of the
amplitude for the emission of two gluons by the proton and their
propagation on $t$ and, yet more importantly, on the gluon
virtualities, the latter being given by the nonperturbative gluon
propagators. We shall see that precisely these two points will have
the main effect on the $t$-dependence of the cross section for our
process.

\subsection{Loop integration}
\label{sec:upper}

Having contracted the tensor for the two gluons coupling to the proton
with the one for their coupling to the $\gamma^\ast \to q \bar{q}$
transition we must perform a loop integration. We label the loop
momentum $l$ in such a way that the first gluon emitted from the
proton carries momentum $ -l +\Delta /2$ and the second gluon $l +
\Delta /2$. Their respective virtualities come out as $(l - \Delta
/2)^2 = (l_T - \Delta_T)^2$ and $(l + \Delta /2)^2 = (l_T +
\Delta_T)^2$, where $l_T$ is the transverse part of $l$ with respect
to $p$ and $q$.  Using the cutting rules we are left with a
two-dimensional loop integral of the form
\begin{equation}
L[f] = \int \frac{d^2 l_T}{\pi} \, [\alpha_s^{(0)}]^2 
D\left[ - (\lt - \dt /2)^2 \right] D\left[ - (\lt + \dt /2)^2\right]
\cdot f
  \label{LoopIntegralsGen}
\end{equation}
with some complicated function $f$ depending on $\lt$, $\dt$ and the
other kinematical variables $\Pt$, $\mq$, $\Mx$, $Q^2$. In particular
$f$ contains quark propagators whose denominators depend on $\lt \cdot
\dt$ and $\lt \cdot \Pt$ so that unlike in the case $t = 0$ we cannot
now perform the integration over the angle of $\lt$ without specifying
a model for the gluon propagator $D$. To obtain a more transparent
representation of the model dependence and to avoid numerical
integrations already at amplitude level we expand those quark
propagators up to second order in $(\lt - \dt /2)$, assuming that both
$|\lt|$ and $|\dt|$ are sufficiently small. The expansion requires
\begin{equation}
\lt^2 \ll \lambda^2 \eqcm \hspace{2em}
| \lt \pdot \Pt | \ll \lambda^2  \eqcm \hspace{2em}
\dt^2 \ll \lambda^2 \eqcm
  \label{ExpansionDb}
\end{equation}
where
\begin{equation}
\lambda^2 = \frac{\Mx^2 /4 - t /4 - \Pl^2}{1 - \beta}
  \label{ScaleGeneral}
\end{equation}
is a generalisation to finite $t$ of the scale \eqref{BartelsScale}.
Note that the first condition is what we used in the approximation
\eqref{LoopApprox} in sec.~\ref{sec:zero}, which was a Taylor
expansion around $\lt^2 / \lambda^2 = 0$. Our calculation will not
give an expression analogous to \eqref{Imn} that does not require the
gluon virtuality to be small compared with the virtuality of the
off-shell quark.

To obtain tractable expressions we also expand denominators in those
terms which depend on the angle $\dq$ between $\Pt$ and $\dt$. The
cross section is then a polynomial in $\cos \dq$ and $\sin \dq$ and
can easily be integrated over $\dq$. The small parameter for these
expansions is again $|\dt |$, more precisely they are valid if
\begin{equation}
| \dt \pdot \Pt | \ll \mq^2 + \Pt^2 + \dt^2 /4  \eqpt
  \label{ExpansionDa}
\end{equation}

\subsection{Integrals over the gluon propagators at finite $t$}
\label{sec:integrals}

After the expansions just described we are left with a limited number
of simple loop integrals. They have the form of $L[f]$ in
\eqref{LoopIntegralsGen} with $f = 1, \: l_T^i, \: l_T^i l_T^j$, where
$i, j = 1, 2$ The corresponding integrands depend only on $\lt$ and
$\dt$ so that the integrals are just functions of $t$. They will turn
out to be crucial quantities in the discussion of our results in
sections~\ref{sec:resultsF} and \ref{sec:resultsP}.

The integral with $f = l_T^i$ is zero because its integrand is odd in
$l_T$, whereas the tensor integral with $f = l_T^i l_T^j$ is related
by rotation invariance to integrals over the scalars $f = \lt^2$ and
$f = (\lt \pdot \dt)^2$. We therefore have three linearly independent
integrals to evaluate and choose the combinations
\begin{eqnarray}
I_0(t) &=& L\left[ 1 \right] \eqcm \nonumber \\
I_1(t) &=& L\left[ \lt^2 \right] \eqcm \nonumber \\
I_2(t) &=& L\left[ 2 (\lt \pdot \dt)^2 / \dt^2 - \lt^2 \right]
 \arreq L\left[ \lt^2 \cos(2 \dl) \right] \eqcm
  \label{BasicIntegrals}
\end{eqnarray}
where $\dl$ is the angle between $\lt$ and $\dt$. At $t = 0$ we have
$I_0(0)= 9 \beta_{0}^{2} /(4 \pi)$ and $I_1(0) = 9 \beta_{0}^{2}
\mu_{0}^{2} /(8 \pi)$ from \eqref{moment1}, \eqref{moment2}, while the
integration over $\dl$ gives $I_2(0) = 0$. The ratio of $I_1(0)$ and
$I_0(0)$ involves the scale $\mu_0^2 \approx 1.2 {\rm \ GeV}^2$, which
is also the characteristic scale for the $t$-dependence of the
integrals in \eqref{BasicIntegrals} since it is the typical scale for
the momentum dependence of $D(l^2)$. We will therefore have two kinds
of corrections to the cross sections and interference terms at zero
$t$:
\begin{enumerate}
\item\label{SmallCorr} corrections in powers of $|t|$ divided by some
  kinematical variable of the $\gp$ reaction, such as $Q^2$, $\Mx^2$,
  or $\lambda^2$ of \eqref{ScaleGeneral}. By assumption these
  kinematical variables are all large compared with $|t|$,
  cf.~\eqref{ExpansionDb}.
\item\label{LargeCorr} corrections which depend on $t$ via $t /
  \mu_0^2$, where $\mu_0^2$ comes from the ratio of the integrals in
  \eqref{BasicIntegrals} or from their variation with $t$. It is
  important that $\mu_0^2$ is {\em not large} compared to typical
  values of $|t| \lsim 1 {\rm \ GeV}^2$.
\end{enumerate}
We have calculated the $\gp$ cross sections and interference terms
keeping the corrections in point \ref{SmallCorr} up to order $t /Q^2$,
whereas no expansion was made in $t /\mu_0^2$, having in mind that we
can have $t /\mu_0^2 = O(1)$.

One can make a more detailed statement about the small-$t$ behaviour
of the integrals \eqref{BasicIntegrals} under the assumption that the
function $D(l^2)$ is sufficiently well behaved to be Taylor expanded.
In \eqref{LoopIntegralsGen} we then can expand $(\lt \mp \dt)^2$ in
the gluon propagators around $\lt^2$. Terms in this expansion that are
odd in $\lt \pdot \dt$ vanish after integration over the angle $\dl$,
so that the integrals have a power expansion in $\dt^2 = |t|\,$:
\begin{eqnarray}
I_0(t) &=& \frac{9 \beta_{0}^{2}}{4 \pi} \left( 1 + c_0^{(1)}
\frac{|t|}{\mu_0^2} + c_0^{(2)} \left(\frac{|t|}{\mu_0^2}\right)^2 +
\ldots \right)  \eqcm \nonumber \\ 
I_1(t) &=& \frac{9 \beta_{0}^{2}}{4 \pi} \, \frac{\mu_0^2}{2} \left( 1
+ c_1^{(1)} \frac{|t|}{\mu_0^2} + c_1^{(2)}
\left(\frac{|t|}{\mu_0^2}\right)^2 + \ldots \right)  
  \eqcm \nonumber \\ 
I_2(t) &=& \frac{9 \beta_{0}^{2}}{4 \pi} \:  |t| c_2^{(0)} \left( 1 +
c_2^{(1)} \frac{|t|}{\mu_0^2} + \ldots \right)   \eqpt
  \label{IntegralsExpand}
\end{eqnarray}
On one hand \eqref{IntegralsExpand} shows that the deviation of these
integrals from their values at $t = 0$ is proportional to $t$ and not
to $\sqrt{-t}$.  Moreover one may get a reasonable description of
their $t$-dependence over a wider range keeping a few terms of this
expansion.

We have evaluated the integrals $I_0(t)$, $I_1(t)$, $I_2(t)$ with the
model \eqref{SpecialGluon} of the gluon propagator. For $n = 4$ we
obtain good quadratic fits of the integrals in the range $|t| = 0$ to
$1.4 {\rm \ GeV }^2$ with
\begin{eqnarray}
c_0^{(1)} &=& - 0.5 , \hspace{1.5em} c_0^{(2)} \arreq 0.12 ,
\hspace{2em} 
c_1^{(1)} \arreq - 0.38 , \hspace{1.5em} c_1^{(2)} \arreq 0.09 ,
\nonumber \\ 
c_2^{(0)} &=& 0.027 , \hspace{1.5em} c_2^{(1)} \arreq - 0.31
  \label{CoeffsGluon}
\end{eqnarray}
and all other coefficients being zero. For $n = \infty$ we have an
exponential propagator $D(- \lt^2) \propto \exp\{- \lt^2 / \mu_{0}^{2}
\}$ and easily find $I_i(t) = \exp\{ - |t| / (2 \mu_{0}^{2}) \} \cdot
I_i(0)$ for $i = 0,1,2$. Comparing the integrals for $n = 4$ and $n =
\infty$ we find that they are almost equal for $I_0$ and that $I_1$ is
smaller for $n = \infty$ than for $n = 4$, the coefficient $c_1^{(1)}$
for $n = \infty$ being $- 0.5$ instead of $- 0.38$. $I_2$ vanishes at
all $t$ for the exponential propagator, for $n = 4$ it is still very
small compared with $|t| \cdot I_0$.

\subsection{Results}
\label{sec:resultsF}

We now present our results for the $\gp$ cross sections and
interference terms of quark-antiquark production at finite $t$, the
$ep$ cross section is obtained from eq.~\eqref{SigmaJetsElectron}. We
first give analytical expressions including the corrections in $t /
\mu_0^2$, but without the corrections of order $t / Q^2$ which are
rather lengthy. The latter will be included in the numerical
discussion below.

To zeroth order in $t / Q^2$ only two linear combinations of the
integrals \eqref{BasicIntegrals} appear, namely
\begin{equation}
K_1(t) = I_1(t) - |t| \, I_0(t) /4  \eqcm \hspace{3em}
K_2(t) = I_2(t) - |t| \, I_0(t) /4  \eqpt
  \label{BasicCombinations}
\end{equation}
From \eqref{IntegralsExpand} we see that the leading term in the
expansion of $K_1(t)$ in $t / \mu_0^2$ is constant, whereas the
leading term for $K_2$ is proportional to $t / \mu_0^2$. We introduce
the abbreviations
\begin{eqnarray}
\sm &=& \frac{\Mx^2 /4 - t /4 - \Pl^2 }{\Mx^2 - t} \limit{t}{0}
\frac{\Pt^2 + \mq^2}{\Mx^2}  \eqcm \nonumber \\
\ib &=& \frac{\Mx^2 /4 - \mq^2 - \Pl^2 }{\Mx^2 /4 - t /4 - \Pl^2 }
\limit{t}{0} \frac{\Pt^2}{\Pt^2 + \mq^2}  \eqcm
  \label{KinAbbrev}
\end{eqnarray}
whose limits for $t \to 0$ are given for easy comparison with our
results in sec.~\ref{sec:zero}, and the variable $\iw = (1 - \beta) \,
\ib$ as a generalisation to finite $t$ of $\iw$ defined in
\eqref{variables}. The result then reads
\begin{eqnarray}
  \lefteqn{\frac{d \sigma^{q\bar{q}}_{mn}}{d \Pl \, d \beta \, d t} =
    } \nonumber \\ && \frac{8}{3} \, \alpha_{\it em} e_q^2 \,
  \frac{\alpha_{s}(\lambda^2)}{\alpha_{s}^{(0)}} \, \tilde{G}^2(t) \,
  \xi^{2 (1 - \alpha_\pom(t))} \, \frac{1 - \beta}{\beta \sqrt{\Mx^2 -
      t} (\Mx^2 + Q^2 - t)^2} \, {\cal S}_{m n}^{q\bar{q}}
  \label{SigmaJetsGamma3}
\end{eqnarray}
with 
\begin{eqnarray}
{\cal S}_{++}^{q\bar{q}} &=& \frac{K_1^2}{\sm^2} \left[ 4 (1 - 2
  \sm) \, \ib \, (1 - 
  \iw)^2 + (1 - \ib) (1 - 2 \iw)^2 \right] +  \nonumber \\
 & & \frac{K_2^2}{\sm^2} \left[ (1 - 2 \sm) \, \ib \, (1 - 2 \iw + 2
  \iw^2) + 2 (1 - \ib) \iw^2 \right] + O(t / Q^2)  \eqcm \nonumber \\
{\cal S}_{00}^{q\bar{q}} &=& \frac{Q^2}{\Mx^2 - t} \, \frac{1}{\sm} \{
  4 K_1^2 \, (1 - 2 \iw)^2 + 8 K_2^2 \, \iw^2 \}  + O(t / Q^2) 
 \eqcm \nonumber \\
{\cal S}_{+-}^{q\bar{q}} &=& \frac{1}{\sm} \left\{8 K_1^2 \, \ib \, (1
  -  \iw)^2 - 4 K_2^2 \, \ib \, \iw (1 - \iw) \right\}  + O(t / Q^2) 
 \eqcm \nonumber \\
{\cal S}_{+0}^{q\bar{q}} &=& - 4 \sqrt{2} \, \frac{Q}{\sqrt{\Mx^2 -
  t}} \, \frac{\Pl \sqrt{\Mx^2 /4 - \mq^2 - \Pl^2}}{\Mx^2 - t} \,
  \frac{1}{\sm^2}  \nonumber \\
 & & \left\{ K_1^2 \, (1 - \iw) (1 - 2 \iw) - K_2^2 \, \iw (1 - 2
  \iw) /2 \right\}  + O(t / Q^2)  \eqpt
  \label{Imn3}
\end{eqnarray}
For $t \to 0$ we recover our previous expressions
\eqref{SigmaJetsGamma}, \eqref{Imn2}.

\begin{figure}
  \begin{center}
    \leavevmode
    \epsfbox{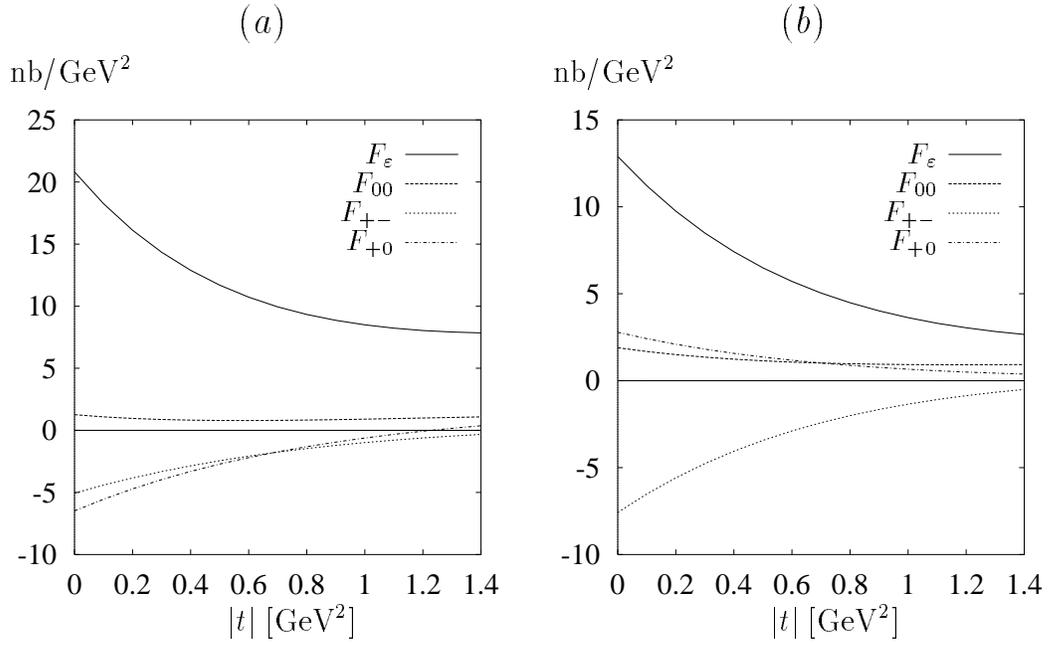}
  \end{center}
  \caption{\label{fig:JetsFinite}Fourier coefficients $F^{FB}_{\eps}$,
    $F^{FB}_{+-}$, $F^{FB}_{+0}$ in the $ep$ cross section and the
    contribution $F^{FB}_{00}$ of longitudinal photons to
    $F^{FB}_{\eps}$. For their definition cf.\ 
    (\protect\ref{FourierJets}). They are summed for $u,d,s$ quarks
    and a global factor $\tilde{G}^2(t) \, \xi^{- 2 \alpha' t}$ is
    taken out in the plot. The results here are obtained with the
    model gluon propagator (\protect\ref{SpecialGluon}) for $n=4$.
    Kinematical variables are $(a)$: $\protect\sqrt{s} = 296 \GeV$, $y
    = 0.5$, $Q^2 = 40 \GeV^2$, $\beta = 1/3$, $|\Pl| \le 4 \GeV$.
    $(b)$: $\protect\sqrt{s}$ and $y$ as before and $Q^2 = 40 \GeV^2$,
    $\beta = 2/3$, $|\Pl| \le 1 \GeV$.}
\end{figure}

Let us now give some numerical examples, obtained with the
parametrisation \eqref{CoeffsGluon} of the integrals $I_i(t)$ which
corresponds to the model gluon propagator \eqref{SpecialGluon} with $n
= 4$. As in sec.~\ref{sec:resultsJ} we change variables from $\Pl$ and
$\phiqq$ to $| \Pl |$ and $\phifb$ and integrate over $|\Pl|$. We
impose an upper cutoff on $|\Pl|$ and sum over the three light quark
flavours $u, d, s$. In fig.~\ref{fig:JetsFinite} we plot the
$t$-dependence of the quantities $F^{FB}_{\eps}$, $F^{FB}_{00}$,
$F^{FB}_{+-}$, $F^{FB}_{+0}$ introduced in \eqref{FourierJets}, but
taking out the squared proton form factor $\tilde{G}^2(t)$ and the
$t$-dependent part $\xi^{- 2 \alpha' t}$ of the Regge power, both of
which give a rather strong suppression of the cross section at $t$
away from zero. As a result of the different $t$-behaviour of
$F^{FB}_{\eps}$, $F^{FB}_{+-}$ and $F^{FB}_{+0}$ the
$\phifb$-dependence of the $ep$ cross section will change with $t$.

\begin{figure}  \begin{center}
    \leavevmode
    \epsfbox{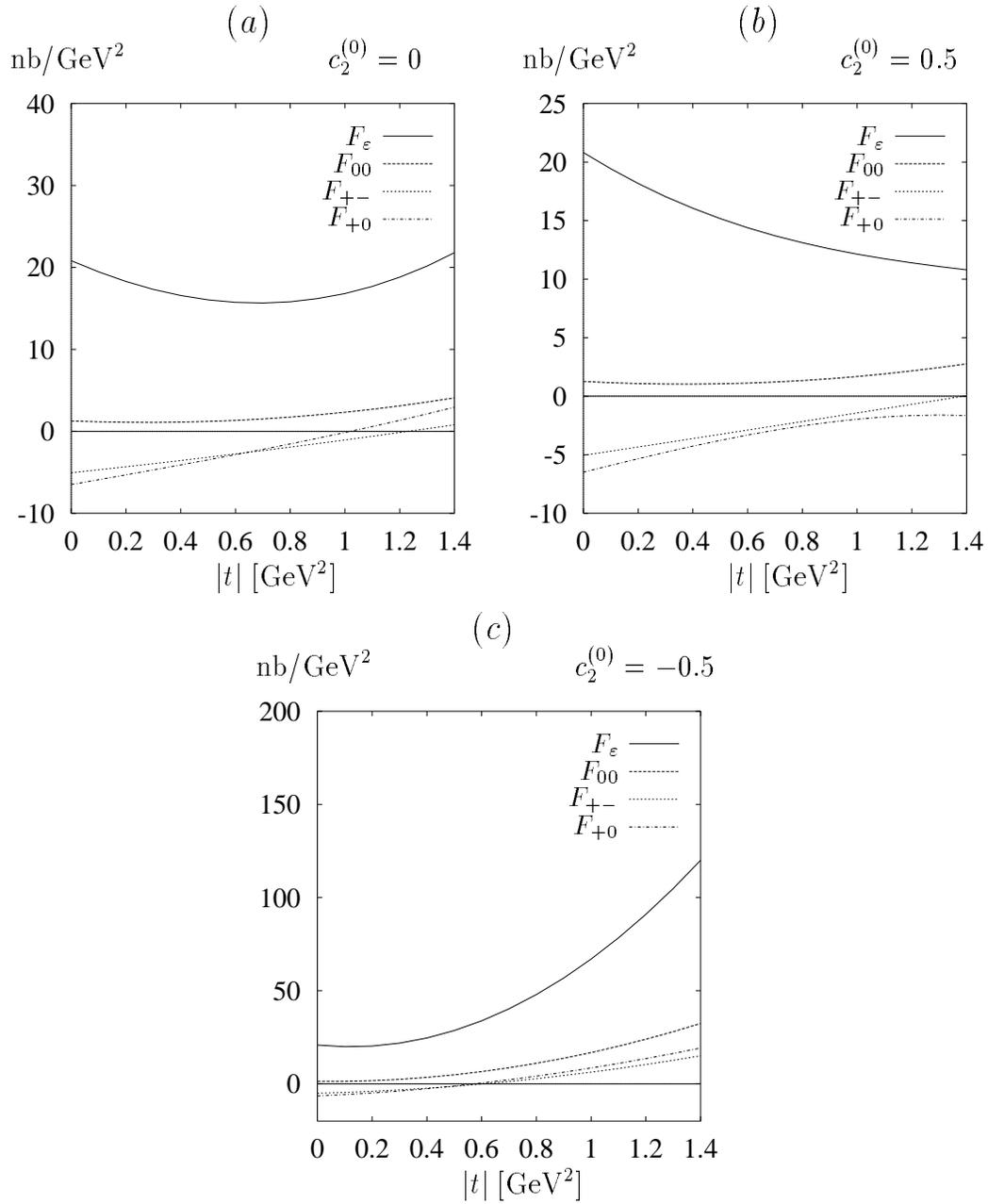}
  \end{center}
  \caption{\label{fig:JetsModel}As fig.~\protect\ref{fig:JetsFinite}
    $(a)$ but with the ansatz (\protect\ref{SimpleExpand}) for the
    integrals over the gluon propagators with different values of
    $c_2^{(0)}$. Remember that a factor $\tilde{G}^2(t) \, \xi^{- 2
      \alpha' t}$ is taken out in the plot; the differential cross
    section does not rise with $|t|$.}
\end{figure}

In order to see to what extent these results depend on the specific
form of the gluon propagator we plot in fig.~\ref{fig:JetsModel} the
same quantities as in fig.~\ref{fig:JetsFinite}, now with the simplest
ansatz for the integrals we can make:
\begin{equation}
I_0(t) = \frac{9 \beta_{0}^{2}}{4 \pi} \eqcm \hspace{3em}
I_1(t) = \frac{9 \beta_{0}^{2}}{4 \pi} \, \frac{\mu_0^2}{2} \eqcm
\hspace{3em} 
I_2(t) = \frac{9 \beta_{0}^{2}}{4 \pi} \:  |t| c_2^{(0)}   \eqcm
\label{SimpleExpand}
\end{equation}
keeping only the lowest order in $t$ of the expansions
\eqref{IntegralsExpand}. The leading coefficient $c_2^{(0)}$ in $I_2$
is not determined from phenomenology as is the case for $I_0$ and
$I_1$, and we take three different values $0$, $0.5$ and $-0.5$. We
see how the variation of $c_2^{(0)}$ modifies the behaviour of the
Fourier coefficients at moderate and large values of $|t|$ quite
drastically; it can for instance lead to a change of sign in the
interference terms at fixed $t$. One would however have to see whether
such large variations of $c_2^{(0)}$ can be obtained with realistic
gluon propagators.


Comparing the plot for $c_2^{(0)} = 0$ and the corresponding one
obtained with our special ansatz for the gluon propagator for which
$c_2^{(0)} \approx 0$ we see that the effect of approximating the
$t$-dependence of the integrals $I_i$ by the leading terms in the
expansions \eqref{IntegralsExpand} is by no means small. This is not
surprising as the leading order approximation is only expected to be
good for $|t| / \mu_0^2 \ll 1$. Notice also that the first order
coefficients $c_0^{(1)}$ and $c_1^{(1)}$ in \eqref{CoeffsGluon} are
rather large.

We have compared the results which include corrections up to order $t
/ Q^2$ with the expressions given in \eqref{Imn3} where only the
$t$-dependence through $t / \mu_0^2$ in the integrals $K_i(t)$ is
kept. In most of parameter space the latter give a very good
approximation, and even for rather small $Q^2$, $\Mx^2$ or rather low
minimum $\Pt^2$ the formulae \eqref{Imn3} give the correct qualitative
features. Apart of course from the squared elastic form factor
$\tilde{G}(t)^2$ and the $t$-dependent pomeron trajectory the main
effect in the $t$-dependence of the $\gp$ cross sections and
interference terms thus turns out to be from the integrals $I_i(t)$,
i.e.\ from the fact that at $t \neq 0$ the two exchanged gluons have
different virtualities. On one hand this means that corrections in $t
/ Q^2$ are less important in the kinematical region we are
investigating. On the other hand the results depend on the details of
the nonperturbative gluon propagator encapsulated in the $I_i(t)$, and
the phenomenological constraints \eqref{moment1}, \eqref{moment2} are
not sufficient to predict the $t$-dependence quantitatively, they only
provide the right order of magnitude and the characteristic scale
$\mu_0^2$.

\section{The azimuthal angle of the scattered proton or proton remnant}
\label{sec:proton}
\setcounter{equation}{0}

In this section we turn our attention to another azimuthal angle in
diffractive processes \eqref{DiffReac}, choosing for the vector $\vp$
\begin{equation}
\vp = p_X  \eqpt
  \label{XDirection}
\end{equation}
$\phiX$ is the azimuthal angle between the lepton and the diffractive
system $X$, i.e.\ $\phiX + \pi$ is the azimuthal angle between the
lepton and the scattered proton or proton remnant.  This angle was
introduced and discussed in \cite{HDCB}. Note that the $\gp$ cross
sections and interference terms $\sigma_{m n}^{(X)}$ introduced there
are integrated over the internal momenta of the system $X$ but not
over $\tilde{p}$. In the notation used in this paper they read
\begin{equation}
\left. \phantom{\frac{1}{\pi}} \sigma_{m n}^{(X)} \, \right|_{\rm
    ref.~[9]} = 
\left. \frac{1}{\pi} \, \frac{\beta}{\xi} \, \frac{d \sigma^X_{m n}}{d
    \beta\, d t} \,  
    \right|_{\rm this\ paper} \eqpt
  \label{translation}
\end{equation}
%
%

$p_X$ becomes collinear with $q$ and $p$ when $|t|$ takes its minimum
value, which is zero in the high energy limit. As we remarked in
sec.~\ref{sec:single} the $\gp$ interference terms must then vanish
for $t \to 0$ because of angular momentum conservation, and the
crucial question we will be concerned with in the following is how
fast they do. To quantify this we normalise the interference terms
with respect to the $\gp$ cross sections and consider the ratios
\begin{eqnarray}
R_{+-} &=& \frac{d \sigma^X_{+-} / (d \beta\, d t)}{d \sigma^X_{++} /
  (d \beta\, d t) + d \sigma^X_{00} / (d \beta\, d t)}  \eqcm 
  \nonumber \\
R_{+0} &=& \frac{d \sigma^X_{+0} / (d \beta\, d t)}{d \sigma^X_{++} /
  (d \beta\, d t) + d \sigma^X_{00} / (d \beta\, d t)}  \eqpt
  \label{RatiosGen}
\end{eqnarray}
If they behave like a (possibly fractional) power of $|t|$ for $t \to
0$ then the scale that compensates $t$ in these dimensionless
quantities determines how large they are at finite $t$.

It was argued in \cite{HDCB} and confirmed by an explicit calculation
in \cite{thesis} that in the phenomenological pomeron model of
Donnachie and Landshoff \cite{l:pomstructure} one has
\begin{equation}
R_{+-} \sim \frac{|t|}{Q^2} \eqcm \hspace{3em}
R_{+0} \sim \frac{\sqrt{|t|}}{Q}  \eqcm
  \label{DLRatios}
\end{equation}
where $Q$ could be replaced by $\Mx$ or some combination of $\Mx$ and
$Q$, the important point is that the scale dividing $t$ is a
kinematical quantity of the $\gp$ subreaction, and therefore rather
large compared with $t$ for the typical values of $t$, $\Mx^2$ and
$Q^2$ in diffractive DIS.

For models that describe diffraction in terms of soft colour
interactions \cite{BuchHeb,Ingelman}, or in models where the QCD
vacuum plays an important role \cite{vacuum}, one can expect a
different behaviour \cite{HDCB}. In these models there is some scale
characteristic of soft physics which could take the place of $Q$ in
the expressions of \eqref{DLRatios}. This would lead to larger
interference terms and thus to a more pronounced $\phiX$-dependence of
the $ep$ cross section. In the LN model the diffractive mechanism is
described by soft gluon exchange; we will see in
sec.~\ref{sec:resultsP} where and when its typical scale $\mu_0$
replaces $Q$ in \eqref{DLRatios}. We remark that the powers of $|t|$
which give the small-$t$ behaviour of $R_{+-}$ and $R_{+0}$ may in
general be different from those in \eqref{DLRatios}.

\subsection{Calculation in the LN model}
\label{sec:protonLN}

We now turn to the predictions of the LN model for the dependence of
the $ep$ cross section on the proton angle. We first have to replace
the general diffractive final state $X$ with a quark-antiquark pair.
This is the lowest order approximation of $X$ at parton level and
should give a reasonable description, except in the region of small
$\beta$, or large diffractive mass $\Mx$, where additional gluon
emission is known to be important.

The calculation of the $\gp$ cross sections and interference terms for
the process \eqref{DiffJets} is essentially the same as the one in
sec.~\ref{sec:finite} with $\qq$ of eq.~\eqref{QuarkDirection}
replaced by $p_X$ in the expressions of the photon polarisation
vectors \eqref{polaris}. We integrate again over the relative
azimuthal angle between ${\bf P}$ and ${\bf p}_X$, but now the
azimuthal angle between ${\bf k}$ and ${\bf p}_X$ is kept fixed, not
the one between ${\bf k}$ and ${\bf P}$.

We have to make an additional restriction on the diffractive final
state, because we need that $\lambda^2$ of eq.~\eqref{BartelsScale},
\eqref{ScaleGeneral} is large for our approximations described in
sec.~\ref{sec:upper} to be valid. Unless we have a large mass $\mq$
for the produced quarks, this means that we must impose a lower cut on
their transverse momentum. 

In \cite{HDCB} it was shown that the $\gp$ cross sections and
interferences have a physical interpretation in terms of the helicity
of the pomeron if one works in the rest frame of the diffractive
system $X$, provided that the selection cuts on the hadronic final
state are invariant when the particle momenta in the system $X$ are
rotated around the photon momentum in this frame while all other
momenta are kept fixed. Nonzero interference terms between photons
with definite helicities in this frame imply that different amounts of
angular momentum along the photon direction are transferred from the
proton. In pomeron language this means that the pomeron can carry
different helicities.

To satisfy the above criterion of rotation invariance we impose a cut
on the transverse quark momentum in the $X$ rest frame and not in the
$\gp$ system. To do this we have to transform the kinematical
quantities introduced in sec.~\ref{sec:JetKin} to the c.m.\ of $X$. We
denote three-momenta with an asterisk there and use a right-handed
coordinate system with the $z$ axis along the photon momentum ${\bf
  q}^\ast$.  Both the transverse and the longitudinal momenta of the
$q \bar{q}$-pair are opposite to each other, not only the longitudinal
ones as in the frame we used in sec.~\ref{sec:JetKin}. Thus ${\bf
  \qq}^\ast = ({\bf P}^\ast_{q} - {\bf P}^\ast_{\bar{q}}) /2$ is equal
to the three-momentum of the quark jet. Instead of $\dq$ and $\Pl$ of
sec.~\ref{sec:JetKin} we use in the $X$ rest system the relative
azimuthal angle $\pp$ between ${\bf \qq}^\ast$ and ${\bf
  \tilde{p}}^\ast$, and the longitudinal momentum $\PlX$ of ${\bf
  \qq}^\ast = {\bf P}^\ast_{q}$. Its kinematical limits are the same
as for $\Pl$. For the transverse component $\PtX$ of ${\bf \qq}^\ast$
in this system we have $\PtPtX = \Mx^2 /4 - \mq^2 - \PlPlX$, compared
with \eqref{QQtransverse}. The relation between the longitudinal and
transverse components of $\qq$ in both frames is as follows:
\begin{eqnarray}
\Pl &=& \frac{1}{\sqrt{1 + 4 \beta^2 t /Q^2}} \,
   \frac{\Mx}{\sqrt{\Mx^2 - t}}   \left( \PlX -
   |{\bf P}_T^\ast| \cos\pp \cdot (1 - 2 \beta) \sqrt{-t} / \Mx
   \right)  \eqcm \nonumber \\
|{\bf P}_T| \cos\dq &=& - \frac{1}{\sqrt{1 + 4 \beta^2 t /Q^2}}
   \left( |{\bf P}_T^\ast| \cos\pp + \PlX  \cdot (1 - 2 \beta)
   \sqrt{-t} / \Mx \right)  \eqcm \nonumber \\
|{\bf P}_T| \sin\dq &=& - |{\bf P}_T^\ast| \sin\pp  \eqpt
\label{conversion}
\end{eqnarray}
Integrating over $\pp$ we obtain the $ep$ cross section differential
in $\phiX$, $x$, $Q^2$, $t$, $\beta$ and $\PlX$. Note that $\PlX$ and
$\pp$ are defined in the $X$ rest frame, but $\phiX$ in the $\gp$
system.

\subsection{Results}
\label{sec:resultsP}

From our master formula \eqref{MasterSimpler} the $\phiX$-dependence
of the $ep$ cross section is given by the analogue of
\eqref{SigmaJetsElectron} in sec.~\ref{sec:zero}, with the
replacements $\phiqq \to \phiX$, $\Pl \to \PlX$ and $d
\sigma^{q\bar{q}}_{m n} \to d \sigma^X_{m n}$. The
transverse-longitudinal interferences are real for the same reason as
in the case of the jet angle. As in sec.~\ref{sec:integrals} we have
calculated the $\gp$ cross sections and interference terms up to order
$t /Q^2$, treating $t / \mu_0^2$ as of order 1. Again we will not give
the analytic expressions of the $O(t / Q^2)$ terms, but use them in
our numerical discussion. We find
\begin{eqnarray}
  \lefteqn{\frac{d \sigma^X_{mn}}{d \PlX \, d \beta \, d t} = }
  \nonumber \\  
&& \frac{8}{3} \, \alpha_{\it em} e_q^2 \,
\frac{\alpha_{s}(\scaleX)}{\alpha_{s}^{(0)}} \,
  \tilde{G}^2(t) \,
\xi^{2 (1 - \alpha_\pom(t))} \, \frac{1 - \beta}{\beta \Mx (\Mx^2 +
  Q^2 - t)^2} \, {\cal S}_{m n}^X
  \label{SigmaXGamma}
\end{eqnarray}
with
\begin{eqnarray}
{\cal S}_{++}^X &=& \frac{K_1^2}{\smsmX} \left[ 4 (1 - 2 \smX) \,
  \ibX (1 - \iwX)^2 + (1 - \ibX) (1 - 2 \iwX)^2 \right] +  
  \nonumber \\
 & &  \frac{K_2^2}{\smsmX} \left[ (1 - 2 \smX) \, \ibX (1 - 2 \iwX + 2
  \iwiwX) + 2 (1 - \ibX) \iwiwX \right] + O(t /Q^2)  \eqcm \nonumber \\
{\cal S}_{00}^X &=& \frac{Q^2}{\Mx^2 - t} \, \frac{1}{\smX} \left\{ 4
  K_1^2  \, (1 - 2 \iwX)^2 + 8 K_2^2 \, \iwiwX \right\} + O(t /Q^2) 
  \eqcm \nonumber \\
{\cal S}_{+-}^X &=&  \frac{8 K_1 K_2}{\smX}  \, \ibX (1 - \iwX)^2 
   + O(t /Q^2)  \eqcm \nonumber \\
{\cal S}_{+0}^X &=& \frac{\sqrt{- 2 t}}{Q} \,
  \frac{Q^2}{\Mx^2 - t} \,
    \frac{1}{\smsmX} \cdot  \nonumber \\
 & & \left\{ K_1^2 \left[ - (1 - \iwX) (1 - 7 \iwX + 8
     \iwiwX) \right. \right. \nonumber \\ 
 & & \hspace{3em} + 4 \smX (1 - \iwX) (1 - 8 \iwX + 10 \iwiwX) 
     \nonumber \\
 & & \hspace{3em} \left. + 2 (1 - 4 \smX) (1 - \beta) (1 - \ibX) (1 -
  6 \iwX + 6 \iwiwX) 
  \right]  \nonumber \\ 
 & & + K_1 K_2  \left[ - (1 - \iwX) (1 - 11 \iwX + 16 \iwiwX) /2
  \right. \nonumber \\ 
 & & \hspace{3em} + 2 \smX (1 - \iwX) (1 - 12 \iwX + 20 \iwiwX) 
     \nonumber \\
 & & \hspace{3em} \left. +  (1 - 4 \smX) (1 - \beta) (1 - \ibX)
  (1 - 10 \iwX + 12 \iwiwX) \right] \nonumber \\
 & & + K_2^2 \left[ \iwX (2 - 9 \iwX + 8 \iwiwX) /2 - 4 \smX \iwX (1 -
  5 \iwX + 5 \iwiwX) \right. \nonumber \\
 & & \hspace{3em} \left. \left. - 2 (1 - 4 \smX) (1 - \beta) (1 -
  \ibX) \iwX ( 1 - 3 \iwX)\phantom{^2}  \right] \right\} \nonumber \\
 & &  + O\left( (-t /Q^2)^{3/2} \right) \eqpt
  \label{ImnX}
\end{eqnarray}
Here we have used the integrals $K_1(t)$, $K_2(t)$ introduced in
\eqref{BasicCombinations}, and the abbreviations
\begin{eqnarray}
\ibX &=& \frac{\PtPtX}{\PtPtX + \mq^2} \eqcm \hspace{3em} 
\iwX \arreq (1 - \beta) \, \ibX  \eqcm \nonumber \\
\smX &=& \frac{\PtPtX + \mq^2}{\Mx^2}  \eqcm \hspace{3em} 
\scaleX \arreq \frac{\PtPtX + \mq^2}{1 - \beta}   \eqpt
  \label{KinAbbrevX}
\end{eqnarray}

A remark is in order on the appearance of the square root $\sqrt{-t}$
in the expression of the transverse-longitudinal interference terms.
One might suspect that there is a contradiction with the analyticity
properties of scattering amplitudes, but this is not so. The point is
that this interference term is multiplied with $\cos\phiX$ in the $ep$
cross section, and that the expression of $\cos\phiX$ in terms of
Mandelstam invariants also involves square roots, so that the
appearance of $\sqrt{-t}$ is a consequence of the kinematical
variables we choose. Put in a different way, the interference terms
can have a dependence on $\sqrt{-t}\,$ through the polarisation
vectors \eqref{polaris}, which contain square roots. A corresponding
remark can be made for the appearance of $\Pl$ in the
transverse-longitudinal interference term corresponding to the jet
angle in \eqref{Imn}, \eqref{Imn3}.

As discussed in sections \ref{sec:integrals} and \ref{sec:resultsF} we
have in the limit of small $t$
\begin{equation}
K_1(t) \limit{t}{0} \frac{9 \beta_{0}^{2}}{4 \pi} \, \frac{\mu_0^2}{2}
\eqcm \hspace{3em} 
K_2(t) \limit{t}{0} \frac{9 \beta_{0}^{2}}{4 \pi}  \: |t|
\left(c_2^{(0)} - 1/4 \right) \eqcm
  \label{CombinationsLimit}
\end{equation}
where $c_2^{(0)}$ is not known from phenomenology and has to be
obtained using a specific ansatz for the nonperturbative gluon
propagator. With \eqref{ImnX} we find for the small-$t$ behaviour of
the interference terms normalised to the $\gp$ cross sections
\begin{equation}
R_{+-} \sim \left(c_2^{(0)} - 1/4 \right) \, \frac{|t|}{\mu_0^2}
\eqcm \hspace{3em} 
R_{+0} \sim \frac{\sqrt{|t|}}{Q}   \eqpt
  \label{LNRatios}
\end{equation}
Note that the behaviour of ${\cal S}_{+-}^X$ in the limit $t \to 0$ is
determined by the coefficient $c_2^{(0)}$, in contrast to the case of
the jet angle investigated in sec.~\ref{sec:resultsF}, where we could
make a parameter-free prediction for this limit. In particular the
sign of the transverse-transverse interference depends on the details
of the gluon propagator. If $c_2^{(0)}$ is close to $1/4$ one even
finds that the $O(t / Q^2)$ terms are dominant and $R_{+-} \sim |t| /
Q^2$. Apart from this caveat the result \eqref{LNRatios} is however
independent of the detailed properties of the gluon propagator. The
scale $\mu_0^2$ comes into play in this model via the nonperturbative
dynamics of the exchanged gluons. Since it only appears squared in the
calculation, cf.~\eqref{IntegralsExpand}, it is clear that it can not
be the scale dividing $\sqrt{-t}$ in $R_{+0}$; there we have the large
kinematical variable $Q$ as in \eqref{DLRatios}.

The situation is however more complicated for $R_{+-}$ than discussed
so far. We will not find ratios $R_{+-}$ of order one as suggested by
\eqref{LNRatios} in our numerical examples. The reason is that ${\cal
  S}_{+-}^X$ (but not ${\cal S}_{+0}^X$) has an additional suppression
compared with the cross section term ${\cal S}_{++}^X$ by a factor of
$\smX \ibX = \PtPtX / \Mx^2$, which can be rather small for the
$\PtPtX / \Mx^2$ where we cut in the case of light quarks, and even
goes down to zero for charm production. After integration over
$\PtPtX$ the first relation of \eqref{LNRatios} is more precisely
\begin{equation}
R_{+-} \sim \left(c_2^{(0)} - 1/4 \right) \, \frac{|t|}{\mu_0^2} 
\cdot  \frac{\PtXcut}{\Mx^2}  \eqcm
  \label{RatioImprovedCut} 
\end{equation}
for light quarks with a momentum cutoff and 
\begin{equation}
R_{+-} \sim \left(c_2^{(0)} - 1/4 \right) \, \frac{|t|}{\mu_0^2} 
\cdot  \frac{\mq^2}{\Mx^2}  \eqcm
  \label{RatioImprovedCharm} 
\end{equation}
for heavy quarks and integration down to $\PtPtX = 0$. In
\eqref{RatioImprovedCharm} $\mq^2$ appears since it is the typical scale
of $\PtPtX$ in the integration if the quark mass is large.

The question arises what one can expect for $R_{+-}$ when there is no
lower cut on $\PtPtX$ in the case of light quarks. After all the main
contribution to $d \sigma^X_{m n} / (d t \, d \beta)$ summed over all
flavours and the full phase space is from light quarks at low
transverse momenta. Because of the approximations of our calculation
we cannot extrapolate \eqref{ImnX} to this region, but we want to give
an educated guess. We have argued in \cite{MD,MD:charm} that with some
caveats the LN model can still be applied to $\gp \to q \bar{q} \, p$
in the limit $\PtPtX + \mq^2 \to 0$. In the results \eqref{Imn} from
our investigation of the jet angle dependence at $t = 0$ we observe
that the suppression of ${\cal S}_{+-}^{q\bar{q}}$ with respect to
${\cal S}_{++}^{q\bar{q}}$ is by a factor $\Pt^2 / \Mx^2$ even in this
limit. Taking this as a guidance for the interference term in our
present problem we expect that a suppression by $\PtPtX / \Mx^2$ of
the differential interference term may persist for very small $\PtPtX
+ \mq^2$, so that $\PtXcut$ in \eqref{RatioImprovedCut} is to be
replaced with some average $\PtPtX$ if we integrate over the full
phase space. Examining the loop integrals $L_i(\Pt^2, w)$ of
\eqref{LoopIntegrals} one further finds that the typical scale for the
$\Pt^2\,$-dependence of the cross section is $\mu_0^2$ in the case
where $\mq^2$ is small and $\beta$ not too close to 1. This leads us
to the guess
\begin{equation}
R_{+-} \sim \left(c_2^{(0)} - 1/4 \right) \, \frac{|t|}{\mu_0^2} 
\cdot  \frac{\mu_0^2}{\Mx^2} \sim \frac{|t|}{Q^2}
  \label{RatioGuess} 
\end{equation}
for the interference term without a cut on $\PtPtX$, which is the
quantity originally discussed in \cite{HDCB}. Notice that the $O(t /
Q^2)$ terms in ${\cal S}_{+-}^X$ now also contribute to the leading
term of $R_{+-}$. The scale $\mu_0^2$ has cancelled and we find the
same behaviour for both $R_{+-}$ and $R_{+0}$ as in the Donnachie
Landshoff model \eqref{DLRatios}. Let us however remark that there one
has $R_{+-} \sim |t| / Q^2$ even with a large cutoff on $\PtPtX$, in
contrast to \eqref{LNRatios}, \eqref{RatioImprovedCut}, so that the
predictions of the two models are by no means identical.

Coming back to what we were able to calculate in the LN model we now
give some numerical illustrations of our results. In analogy to
sec.~\ref{sec:resultsF} we integrate over $\PlX$ and plot the Fourier
coefficients $F^X_{\eps}$, $F^X_{+-}$, $F^X_{+0}$ with a global factor
$\tilde{G}^2(t) \, \xi^{- 2 \alpha' t}$ taken out. They are defined
like $F^{\it FB}_{\eps}$, $F^{\it FB}_{+-}$, $F^{\it FB}_{+0}$ in
\eqref{FourierJets} with the superscript ${\it FB}$ replaced by $X$
and thus appear in $d \sigma(e p \to e p \, q \bar{q}) /(d \phiX \, d
x\, d Q^2 \, d \beta \, dt)$ in a manner analogous to
\eqref{FourierInSigma}. We either sum over the three light flavours
$u,d,s$ for the produced quarks, with a minimum $\PtPtX$ so that our
calculation is valid, or we consider produced charm quarks for which
we can integrate over the full kinematical range of $\PlX$.

\begin{figure}
  \begin{center}
    \leavevmode
    \epsfxsize=0.95\textwidth  \epsfbox[90 149 487 616]{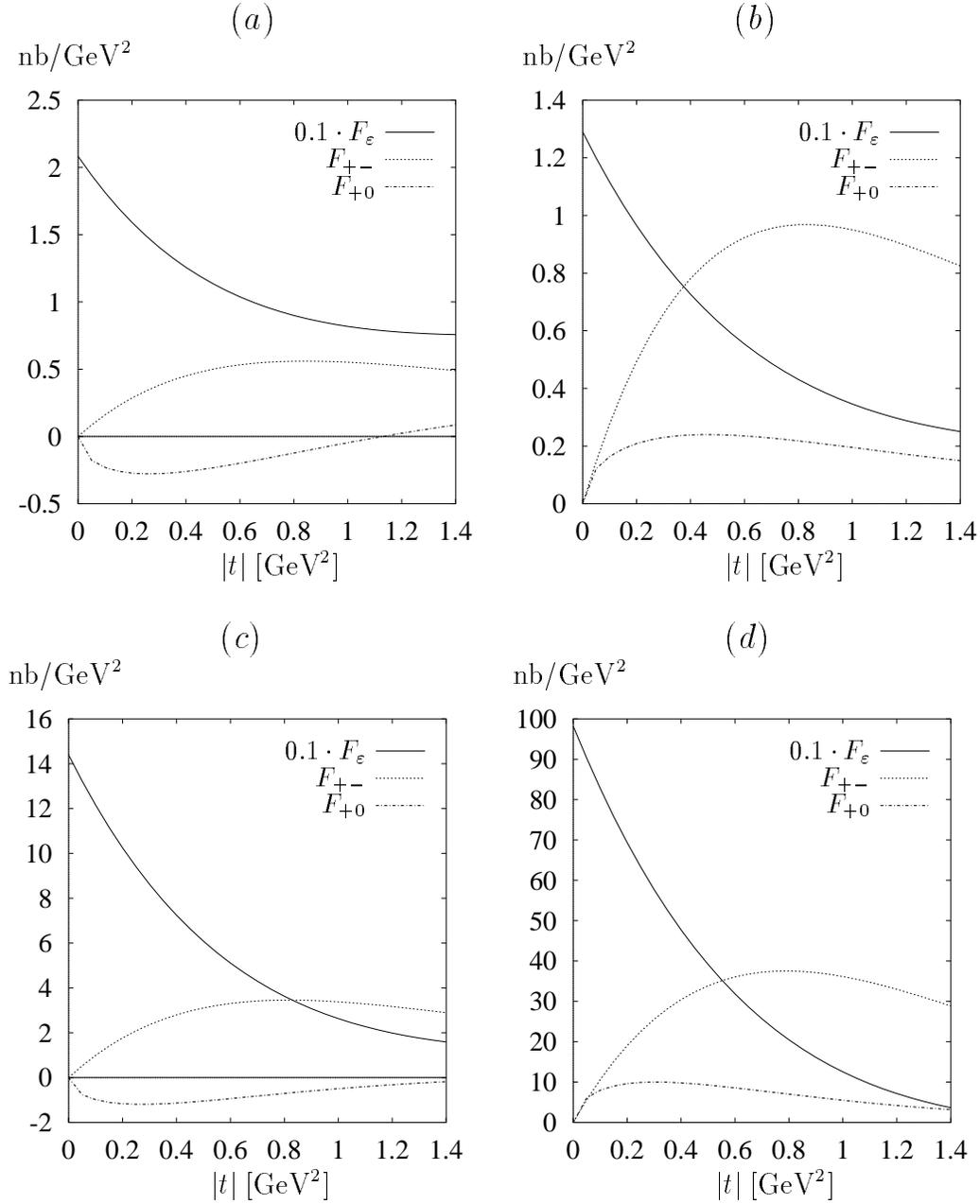}
  \end{center}
  \caption{\label{fig:FourierProt}Fourier coefficients $F^X_{\eps}$,
    $F^X_{+-}$, $F^X_{+0}$ in the $ep$ cross section, defined in
    analogy to (\protect\ref{FourierJets}), with a global factor
    $\tilde{G}^2(t) \, \xi^{- 2 \alpha' t}$ taken out. Note that
    $F^X_{\eps}$ is scaled down by a factor 10. $(a)$ and $(b)$ are
    summed for $u,d,s$ quarks with a lower cutoff $\PtPtX \ge 4
    \GeV^2$, $(c)$ and $(d)$ are for charm quarks without a cut on
    $\PtPtX$. Kinematical variables are $\protect\sqrt{s} = 296 \GeV$,
    $y = 0.5$ in all cases and $(a)$: $Q^2 = 40 \GeV^2$, $\beta =
    1/3$.  $(b)$: $Q^2 = 40 \GeV^2$, $\beta = 2/3$. $(c)$: $Q^2 = 25
    \GeV^2$, $\beta = 1/3$.  $(d)$: $Q^2 = 6.25 \GeV^2$, $\beta =
    1/3$. The curves are obtained with the model gluon propagator
    (\protect\ref{SpecialGluon}) with $n=4$.}
\end{figure}

\begin{figure}
  \begin{center}
    \leavevmode
    \epsfbox{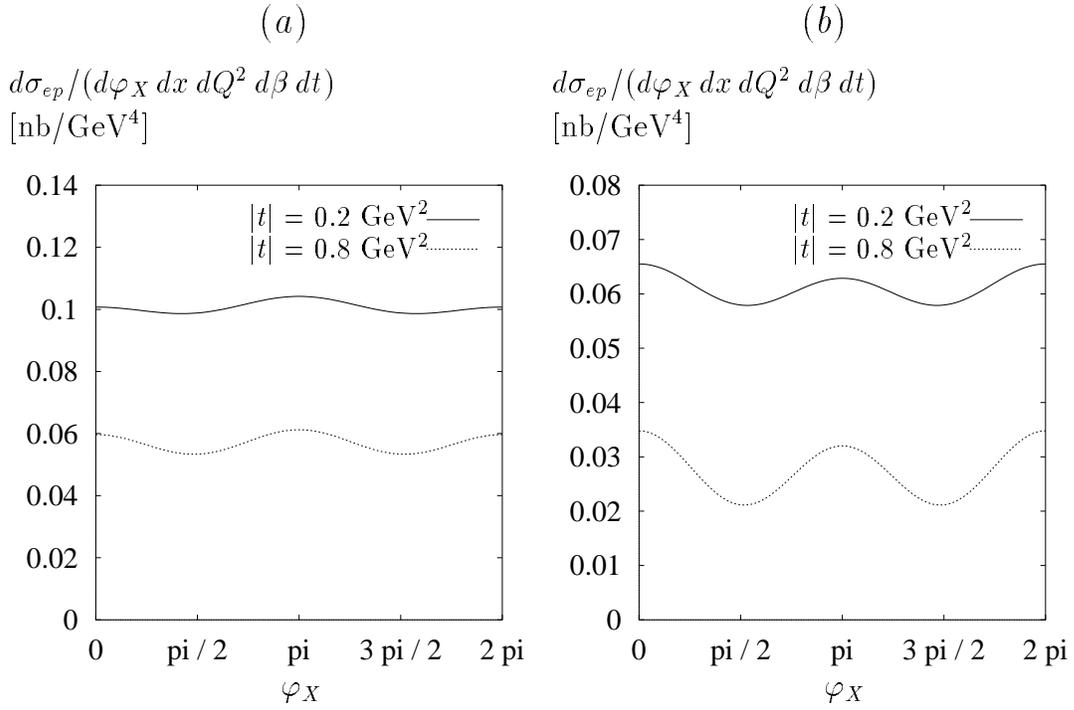}
  \end{center}
  \caption{\label{fig:AngleProt}$(a)$: Dependence on $\phiX$ of $d
    \sigma(ep \to ep \, q \bar{q}) / (d \phiX \, d x\, d Q^2 \, d
    \beta \, dt)$, obtained from the Fourier coefficients in
    fig.~\protect\ref{fig:FourierProt} $(a)$ for $|t| = 0.2 \GeV^2$
    and $|t| = 0.8 \GeV^2$. $(b)$: The same for the Fourier
    coefficients from fig.~\protect\ref{fig:FourierProt} $(b)$ }
\end{figure}

The $t$-dependence of the Fourier coefficients for the model
propagator \eqref{SpecialGluon} with $n=4$ is shown in
fig.~\ref{fig:FourierProt} for different values of the free
kinematical parameters. We observe that the transverse-transverse
interference is usually larger than the transverse-longitudinal one.
Fig.~\ref{fig:AngleProt} shows an example of the $\phiX$-dependence of
the $ep$ cross section at two different values of $t$. In case $(b)$
the distribution is clearly not flat, although the effect is not very
large, whereas in case $(a)$ almost no $\phiX$-dependence can be seen.
This illustrates how the transverse-transverse interference is
affected by the parameter $\PtXcut / \Mx^2$, which is $1/20$ in case
$(a)$ and $1/5$ in case $(b)$.

\begin{figure}
  \begin{center}
    \leavevmode
    \epsfbox{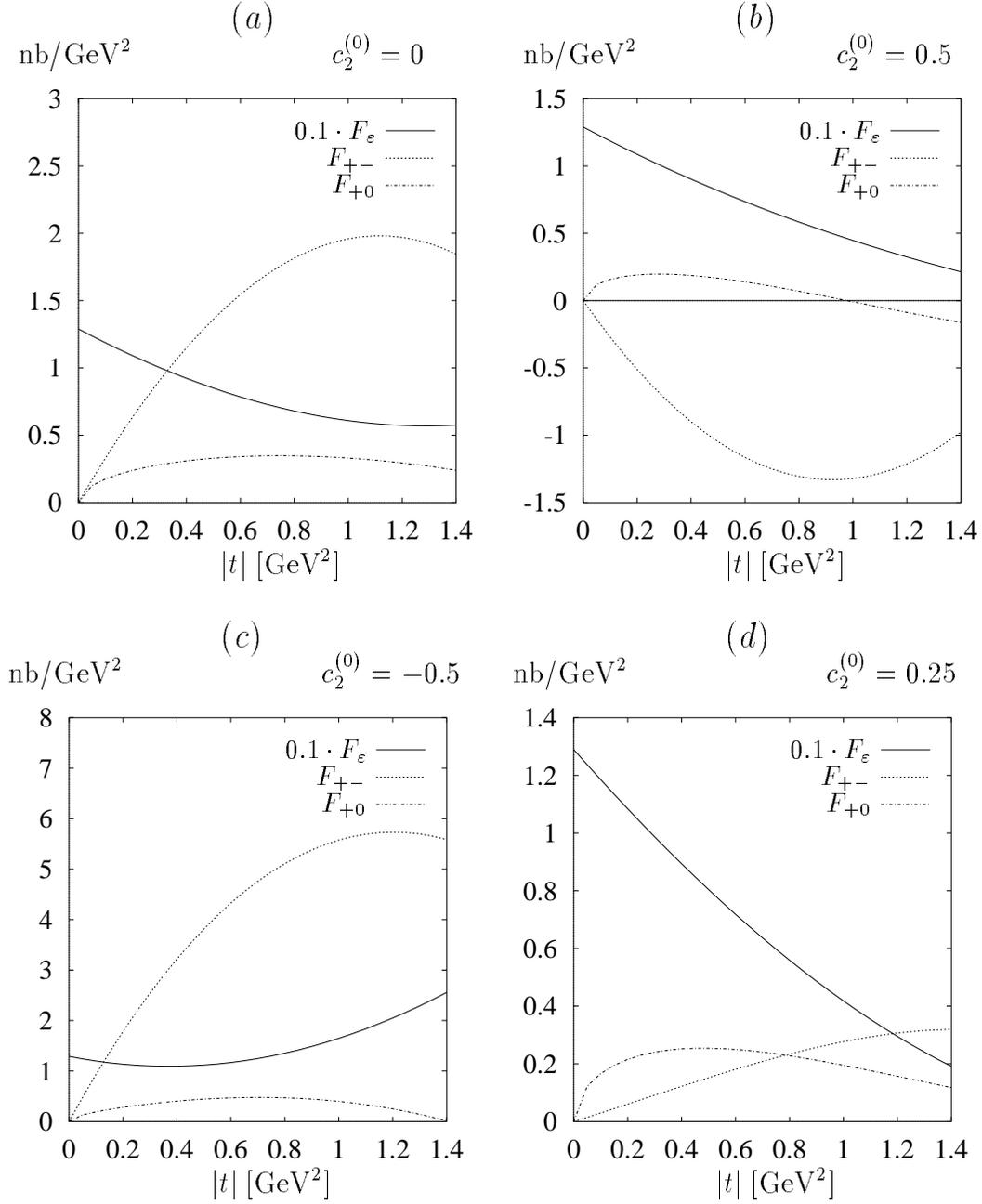}
  \end{center}
  \caption{\label{fig:FourierProtModel}As
    fig.~\protect\ref{fig:FourierProt} $(b)$, but with the ansatz
    (\protect\ref{SimpleExpand}) for the integrals over the gluon
    propagators with different values of $c_2^{(0)}$.}
\end{figure}

To assess the model dependence of our prediction we also evaluated the
Fourier coefficients taking the simple ansatz \eqref{SimpleExpand} for
the integrals over the gluon propagators, with different values for
the coefficient $c_2^{(0)}$. They are shown in
fig.~\ref{fig:FourierProtModel} for the case of light quarks, the
effects for charm are similar. As in sec.~\ref{sec:resultsF} the
results, especially at large $|t|$, change considerably with
$c_2^{(0)}$. In particular the sign of the transverse-transverse
interference term is different for $c_2^{(0)}$ below or above $1/4$,
and for $c_2^{(0)} = 1/4$ this term is very small, as discussed above.
We repeat that one would have to see whether realistic gluon
propagators give values of $c_2^{(0)}$ as far away from zero as the
ones taken in fig.~\ref{fig:FourierProtModel}.


We find that the leading order expressions of ${\cal S}_{++}^X$,
${\cal S}_{00}^X$ and ${\cal S}_{+-}^X$ in \eqref{ImnX} give a rather
good approximation of what is obtained by including terms of $O(t /
Q^2)$, except of course for ${\cal S}_{+-}^X$ if $c_2^{(0)} = 1/4$. As
in sec.~\ref{sec:resultsF} this means that the main effect comes from
terms depending on $t / \mu_0^2$, whereas corrections in $t / Q^2$ are
relatively small. Terms in $t / \mu_0^2$ are also essential to
describe the $t$-dependence of ${\cal S}_{+0}^X$. Although its order
of magnitude at small $t$ is given by $\sqrt{-t} /Q$, a mere square
root dependence on $|t|$ for the Fourier coefficient $F^X_{+0}$ in
figures \ref{fig:FourierProt} and \ref{fig:FourierProtModel} is
clearly not a good approximation unless $t$ is very small.

To conclude this section we remark on the possibility to constrain the
cross section for longitudinal photons from the measurement of the
$\phiX$-dependence using the method described in
sec.~\ref{sec:bounds}. With the results we obtain in the LN model the
bounds on $d \sigma_{00} / (d t \, d \beta)$ would not be stringent at
all, and be far away from its actual value. This is because we find
the interference $d \sigma_{+0} / (d t \, d \beta)$, whose size is
crucial to obtain good constraints, to be of order $\sqrt{|t|} / Q$.
From \eqref{constraint3} we can see that the lower bound on $d
\sigma_{00} / (d t \, d \beta)$ then vanishes like $|t| / Q^2$ for
small $t$, whereas $d \sigma_{00} / (d t \, d \beta)$ itself does not
become small in this limit. This is different from the situation we
found for the jet angle in sec.~\ref{sec:resultsJ}. One would expect
better bounds if the ratio of the longitudinal-transverse interference
and the $\gp$ cross sections at small $t$ were dominated by a hadronic
scale instead of $Q$ or $\Mx$. This might happen in other models of
diffraction where soft dynamics is important.

\section{Summary}
\label{sec:sum}
\setcounter{equation}{0}

In this paper we investigated correlations between azimuthal angles in
deep inelastic $ep$ diffraction, using the one-photon approximation.
We first derived the general expression for the dependence of the $ep$
cross section on a suitably defined azimuthal angle between the lepton
plane and a direction in the hadronic final state in terms of cross
sections and interference terms of the $\gp$ collision for different
photon helicities. This was a direct generalisation of the work in
\cite{HDCB}. We showed that those terms in the cross section that
depend on the helicity of the lepton beam are sensitive to a
polarisation dependence of the phases in the $\gp$ amplitudes for
linearly polarised photons. From the angular dependence of the $ep$
cross section one can obtain bounds on the differential or integrated
cross section for longitudinal photons, without having to vary $y$ as
it is needed for its direct measurement. How stringent these bounds
are depends on the size of the interference term between longitudinal
and transverse polarisations and thus on the choice of azimuthal angle
and on the region of phase space considered.

We have investigated the dependence on the azimuthal jet angle
predicted by the LN model for the parton level reaction $ep \to ep + q
\bar{q}$ at large transverse momentum of the $q \bar{q}$-pair, which
at hadron level describes a pair of jets that carries the entire
four-momentum of the diffractive final state. The size of the
interference terms is found to be controlled by the quantity $\Pt^2 /
\Mx^2$. The sign of the transverse-longitudinal interference depends
on $\beta$. Since this interference is less strongly suppressed than
the longitudinal cross section, it may offer an opportunity to observe
the zero of the longitudinal amplitude at certain values of $\Pt^2$
and $\beta$ which is characteristic of the two-gluon exchange
mechanism.  The bounds on the longitudinal cross section obtained from
the azimuthal dependence might be quite useful, at small $\Pt^2
/\Mx^2$ at least the lower bound comes out quite close to its actual
value which is also small in this kinematical region. We suggest that
the use of an azimuthal angle defined from an event shape variable
like the thrust axis in the diffractive final state would allow to
extend this method to a wider class of final states, in particular it
would allow to go to smaller values of $\Pt^2$ than those needed for
jet algorithms and thus to increase the total rate in the analysis.

The cross section for $ep \to ep + q \bar{q}$ was then calculated at
finite $t$ with the approximations $\dt^2 \ll \lambda^2$ and $\lt^2
\ll \lambda^2$, for the definition of $\lambda^2$
cf.~\eqref{BartelsScale}, \eqref{ScaleGeneral}. Its region of validity
is therefore the production of jets or heavy flavours where
$\lambda^2$ is sufficiently large. The result then involves three
$t$-dependent integrals \eqref{BasicIntegrals} with two gluon
propagators at different virtualities. The relevant scale for the
$t$-behaviour of these integrals is $\mu_0^2 \approx 1.2 \GeV^2$. The
limit $t \to 0$ for two of them is known from phenomenology, for the
rest one has to resort to specific model propagators.

Applying this calculation to the dependence on the azimuthal jet angle
we find that apart from the dominating effect of the proton form
factor $\tilde{G}(t)$ and the pomeron trajectory $\alpha_\pom(t)$ the
$t$-dependence of the $\gp$ cross sections and interference terms is
controlled by corrections in $t / \mu_0^2$ coming from the gluon
propagators, corrections in $t$ divided by a large kinematical scale
of the transition $\gamma^\ast \to q \bar{q}$ are much smaller. As a
consequence the quantitative features of the results depend on the
choice of gluon propagator.  Using the model propagator
\eqref{SpecialGluon} we typically find that the sum of the transverse
and longitudinal cross sections decreases by a factor around 2 between
$|t| = 0$ and $1.4 \GeV^2$ when the strong suppression from
$\tilde{G}(t)$ and the pomeron trajectory is taken out. The absolute
size of the interference terms tends to decrease with $|t|$ and one
can even have a change of their signs.

Another important azimuthal angle is that of the scattered proton or
proton remnant. It was shown in \cite{HDCB} that its measurement can
give information on the helicity structure of the pomeron. We have
investigated its distribution in the LN model, but due to our
approximations had to restrict ourselves to a $q \bar{q}$ final state
with large transverse quark momentum $\PtX$ in the $q \bar{q}$ rest
frame, or with a large quark mass. Like for the jet angle we find that
the $t$-dependence of the cross sections and interference terms is
controlled by the scale $\mu_0^2$, and that the results depend rather
strongly on the integrals over the gluon propagators. The order of
magnitude of the transverse-longitudinal interference is given by
$\sqrt{-t} /Q$. The ratio between the interference of the two
transverse polarisations and the transverse cross section goes like $t
/ \mu_0^2$ which can be large, but it is suppressed by an additional
factor $\PtPtX / \Mx^2$ so that this interference is small at low
$\PtPtX$. Unfortunately we cannot take the limit $\PtPtX \to 0$ for
light quarks in our calculation but our guess is that the
$\PtPtX\,$-integrated transverse-transverse interference will be
suppressed by $t / Q^2$ compared with the transverse cross section,
which would lead to a rather flat angular dependence in the $ep$
spectrum. According to the discussion in \cite{HDCB} the helicity of
the LN pomeron is then dominated by one value in the inclusive
diffractive process, whereas several helicities are important when
there is a high transverse momentum or mass scale in the diffractive
final state. Notice that a system of two gluons we use to model the
pomeron can in principle transfer any integer value of angular
momentum through its orbital motion.

Our finding that finite-$t$ effects are rather sensitive to the
nonperturbative gluon dynamics in the LN model suggests that they may
come out quite different in other models of diffraction and could thus
be a useful probe of the mechanisms at work in diffractive physics.

\subsection*{Acknowledgements}

I wish to thank O. Nachtmann and P. V. Landshoff for numerous
discussions, and O. Nachtmann and B. Pire for valuable remarks on the
manuscript. I gratefully acknowledge conversations with T. Arens,
J. Bartels, H. Jung, H. Lotter, N. Pavel and H. Pirner.

This work was supported by the ARC Programme of the British Council
and the German Academic Exchange Service, grant 313-ARC-VIII-VO/scu,
and by the EU Programme ``Human Capital and Mobility'', Network
``Physics at High Energy Colliders'', Contracts CHRX-CT93-0357 (DG 12
COMA) and ERBCHBI-CT94-1342. It was also supported in part PPARC.


\newpage

\end{document}